\documentclass[twoside,11pt]{article}

\usepackage[abbrvbib, preprint]{jmlr2e}

\usepackage{amsmath,amsfonts,setspace,natbib,amssymb,ifthen,url,array,color,hyperref,multirow}
\newcommand{\x}{\mathbf{x}}

\newcommand{\E}{\mathbb{E}}
\newcommand{\distance}{\mathrm{distance}}
\newcommand{\MG}{\mathrm{MG}}

\newcommand{\malts}{\textsc{MALTS}}
\newcommand{\dis}{\mathbf{d}}
\newtheorem{define}{Definition}

\DeclareMathOperator*{\argmin}{arg\,min}
              
\newcommand{\indep}{\rotatebox[origin=c]{90}{$\models$}}

\definecolor{brown}{rgb}{0.8,0.1,0.1}
\definecolor{BROWN}{rgb}{0.8,0.1,0.1}
\definecolor{red}{rgb}{1,0,0}
\definecolor{RED}{rgb}{1,0,0}

\definecolor{dartmouthgreen}{rgb}{0.05, 0.5, 0.06}
\definecolor{burntumber}{rgb}{0.54, 0.2, 0.14}
\definecolor{BURNTUMBER}{rgb}{0.54, 0.2, 0.14}

\bibpunct[, ]{(}{)}{,}{a}{}{,}

\usepackage{bbm}
\usepackage{lscape}

\usepackage{lastpage}
\jmlrheading{23}{2022}{1-\pageref{LastPage}}{1/21; Revised
7/22}{8/22}{21-0053}{Harsh Parikh, Cynthia Rudin and Alexander Volfovsky}

\ShortHeadings{MALTS}{Parikh, Rudin and Volfovsky}

\firstpageno{1}

\title{\textsc{MALTS: Matching After Learning to Stretch}}
	
 \author{\name Harsh Parikh \email           harsh.parikh@duke.edu \\
        \addr Department of Computer Science\\
        Duke University\\
        Durham, NC 27708-0129, USA.
        \AND
        \name Cynthia Rudin \email cynthia@cs.duke.edu \\
        \addr Department of Computer Science\\
        Duke University\\
        Durham, NC 27708-0129, USA.
        \AND
        \name Alexander Volfovsky \email alexander.volfovsky@duke.edu\\
        \addr Department of Statistical Science\\
        Duke University\\
        Durham, NC 27710, USA.
     }

\begin{document}
	\editor{Russ Greiner}
  	\maketitle
    \begin{abstract}
		We introduce a flexible framework that produces high-quality almost-exact matches for causal inference. Most prior work in matching uses ad-hoc distance metrics, often leading to poor quality matches, particularly when there are irrelevant covariates. In this work, we learn an interpretable distance metric for matching, which leads to substantially higher quality matches. The learned distance metric stretches the covariate space according to each covariate's contribution to outcome prediction: this stretching means that mismatches on important covariates carry a larger penalty than mismatches on irrelevant covariates.
       Our ability to learn flexible distance metrics leads to matches that are interpretable and useful for the estimation of conditional average treatment effects.
	\end{abstract}

	\allowdisplaybreaks
	\section{Introduction}
Matching methods are used throughout the social and health sciences to make causal conclusions where access to randomized trials is scarce but observational data are widely available. 
Matching methods construct sets of similar individuals, some of whom select into treatment and some of whom select into control, allowing for direct comparison of outcomes between the samples from these populations. These methods are particularly interpretable since they allow fine-grained troubleshooting of the data. 
For instance, examining a matched group of patients through chart review of their medical data and doctors' notes may allow an analyst to determine whether the matched groups are indeed trustworthy, and if not, determine what other factors should be included in the analysis.
Having high-quality matches also allows the user to estimate nonlinear treatment effects with lower bias than parametric approaches.

As a concrete example of the importance of match group quality, Table~\ref{tab:ex_mg} presents a series of matched groups from the Lalonde dataset \citep{lalonde, dehejia_wahba_nonexp}. A simple visual inspection of the matched groups produced by standard-bearer methods like propensity score matching and prognostic score matching reveals that the units being considered similar by these methods are not similar on underlying covariates. On the other hand, the matches generated by our proposed method are qualitatively (and quantitatively) better. \emph{The quality of the matches is our main consideration in this work.} 

\begin{table}[h]
\centering
\caption{Example control units in a matched group for a treated unit using (a) our approach (MALTS), (b) prognostic score \citep{hansen2008prognostic}, and (c) propensity score matching \citep{rosenbaum1983central} for a query unit in the Lalonde dataset (top rows). Our method matched closely on covariates -- age, education, whether the person had an academic degree, and income in 1975. In contrast, prognostic and propensity scores did not match closely on these factors.
}
\label{tab:ex_mg}
\resizebox{\textwidth}{!}{\begin{tabular}{l|c|rrrrrrr|r}
\hline
\multicolumn{1}{c}{} & \multicolumn{1}{c|}{\textbf{Treatment}} & \multicolumn{7}{c|}{\textbf{Covariates}} & \multicolumn{1}{c}{\textbf{Outcome}} \\ \hline
\multicolumn{1}{l|}{\textit{\textbf{Unit ID}}} & \multicolumn{1}{r|}{\textbf{Treated}} & \textbf{Age} & \textbf{Education} & \textbf{Black} & \textbf{Hispanic} & \textbf{Married} & \textbf{No-Degree} & \multicolumn{1}{r|}{\textbf{Income-1975}} & \textbf{Income-1978} \\ \hline
\textit{Query: 1} & \multicolumn{1}{c|}{Yes} & 22 & 9 & No & Yes & No & Yes & \$0 & \$3596 \\ \hline
\multicolumn{10}{c}{\textbf{(a) Our Approach (MALTS)}} \\ \hline
330 & \multicolumn{1}{c|}{No} & 22 & 8 & No & Yes & No & Yes & \$0 & \$9921 \\
299 & \multicolumn{1}{c|}{No} & 22 & 9 & \textbf{Yes} & \textbf{No} & No & Yes & \$0 & \$0 \\
416 & \multicolumn{1}{c|}{No} & 22 & 9 & \textbf{Yes} & \textbf{No} & No & Yes & \$0 & \$12898 \\
\hline
\multicolumn{10}{c}{\textbf{(b) Prognostic Scores}} \\ \hline
338 & \multicolumn{1}{c|}{No} & \textbf{44} & 9 & \textbf{Yes} & \textbf{No} & No & Yes & \$0 & \$9722 \\
340 & \multicolumn{1}{c|}{No} & 22 & \textbf{12} & \textbf{Yes} & \textbf{No} & No & \textbf{No} & \textbf{\$532} & \$1333 \\
355 & \multicolumn{1}{c|}{No} & 18 & 10 & No & Yes & No & Yes & \$0 & \$1859 \\ \hline
\multicolumn{10}{c}{\textbf{(c) Propensity Scores}} \\ \hline
451 & \multicolumn{1}{c|}{No} & 22 & 8 & \textbf{Yes} & \textbf{No} & No & Yes & \$0 & \$1391 \\
330 & \multicolumn{1}{c|}{No} & 22 & 8 & No & Yes & No & Yes & \$0 & \$9921 \\
407 & \multicolumn{1}{c|}{No} & 20 & \textbf{12} & \textbf{Yes} & \textbf{No} & No & \textbf{No} & \textbf{\$1371} & \$20893 \\ \hline
\end{tabular}}
\end{table}

Typically, matching methods place units that are close together into the same matched group, where closeness is measured in terms of a pre-defined distance (e.g., exact, coarsened exact, Euclidean, etc.), while maintaining balance constraints between treatment and control units. Despite its merits, this classical paradigm has flaws, namely that it relies heavily on a prespecified distance metric. The distance metric cannot be determined without an understanding of the importance of the variables;
for instance, the quality of matches for any prespecified distance that weighs all covariates equally will degrade as the number of irrelevant covariates increases. 
This is true irrespective of the matching methodology employed.
This issue has previously been referred to as the toenail problem \citep{wang2017flame, DiengEtAl2018}, where the inclusion of irrelevant covariates (like ``toenail length'') with nonzero weights can worsen the metric for matching. A related concern is that the covariates may be scaled differently, where a given distance along one covariate has a different impact than the same distance along a different covariate; in this case, if the scaling or weights on the covariates are chosen poorly, the total distance metric can inadvertently be determined by less relevant covariates, again leading to lower quality matches.

Ideally, the distance metric would focus on important covariates that significantly contribute to the outcome, so that after matching, treatment effect estimates computed using the matched groups would be accurate. If the researcher knows how to choose the distance metric so that it yields accurate treatment effect estimates, it would solve the problem. However, there is no reason to believe that this is achievable in complex high-dimensional data settings. Producing high dimensional functions to characterize data is a task at which humans are not naturally adept.

In this work, we propose a framework for matching where an interpretable distance measure between matched units is learned from a training set. As long as the distance metric generalizes from the training set to the full sample, we are able to compute high-quality matches and accurate estimates of conditional average treatment effects (CATEs) within the matched groups. One can use any form of distance metric to train, and in this work, we focus on exact matching for discrete variables and generalized Mahalanobis distances for continuous variables. By definition, the generalized Mahalanobis distance is determined by a matrix. If the matrix is diagonal, the distance calculation represents a stretch for each covariate. Irrelevant covariates will be compressed so that their values are always effectively zero. Highly relevant covariates will be stretched so that for two units to be considered a match, they must have very similar values for those covariates. In this way, diagonal matrices lead to very interpretable distance metrics. If the Mahalanobis distance matrix is not constrained to be diagonal, then it induces a stretch and rotation, leading to more flexible but less interpretable notions of distance. 

\begin{figure}
    \centering
    \includegraphics[width=0.9\textwidth]{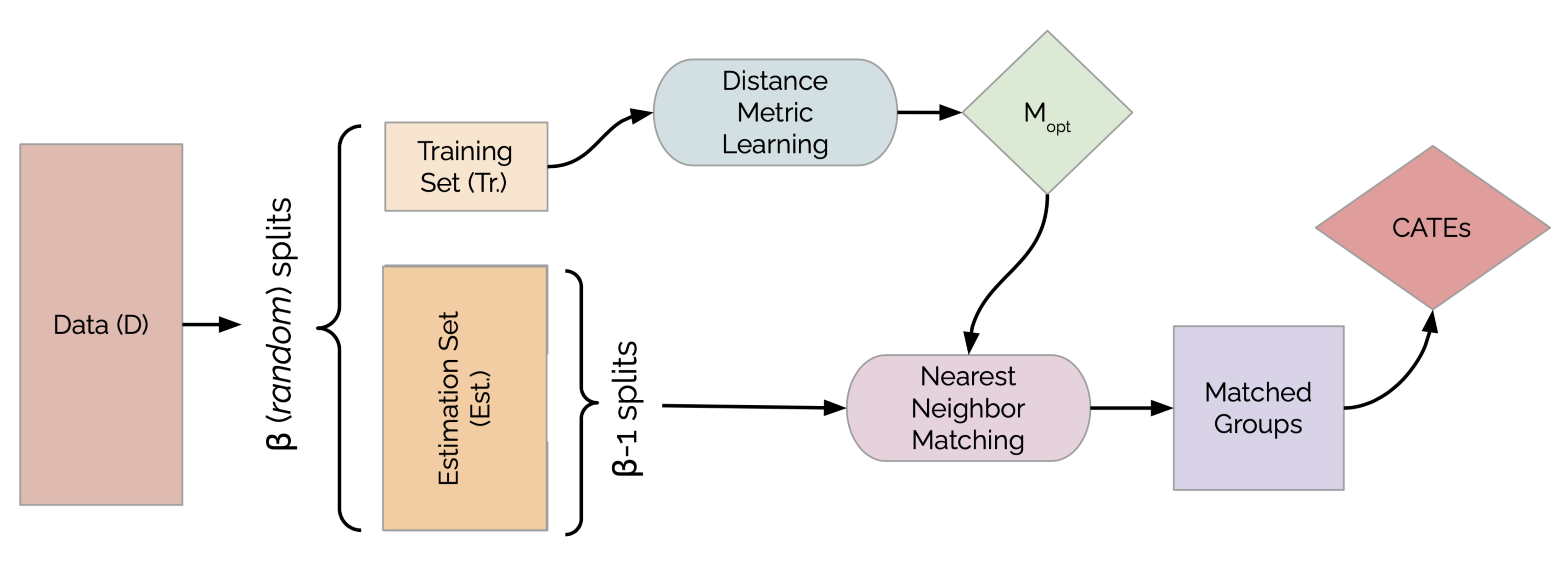}
    \caption{Schematic drawing of MALTS algorithm. The algorithm splits the data into random subsets and uses one of the subsets (training set) to learn a distance metric. It performs matching on the rest of the units (estimation set) using the learned distance metric to produce tightly matched groups and estimate conditional average treatment effects.}
    \label{fig:malts_framework}
\end{figure}

The new framework is called Learning-to-Match, and the algorithm introduced in this work is called Matching After Learning to Stretch (MALTS).  Figure~\ref{fig:malts_framework} shows the main steps of MALTS, which are: divide the data into training and estimation sets, learn the distance metric on the training set, use the learned distance metric to perform nearest neigbhor matching on estimation set, and use those matched groups to estimate conditional average treatment effects (CATEs). 
We tested MALTS against several other matching methods in simulation studies (Section~\ref{sec:Experiments}), where ground truth CATEs are known. In these experiments, MALTS consistently achieves substantially better results than other matching methods including Genmatch, propensity score matching, and prognostic score matching for estimating CATEs. Even though our method is heavily constrained to produce interpretable matches, it performs at the same level as non-matching methods that are designed to fit extremely flexible but uninterpretable models directly to the response surface. 

In Section~\ref{sec:framework}, we introduce the learning-to-match framework and show that under a choice of smooth distance metric (Definition~\ref{def:smoothdis}) we can estimate conditional average treatment effects accurately with high probability.  Section~\ref{sec:method} discusses MALTS' optimization set up and training procedure that learns a smooth distance metric. In Section~\ref{sec:theory}, we prove that the distance metric learned by MALTS is multi-robust (Definition~\ref{def:robust}) and generalizable (Definition~\ref{def:general}). Thus, the distance metric estimated by MALTS' framework facilitates the correct estimates of CATEs under SUTVA and positivity assumptions. 

 	\section{Related work}

Since the 1970's, the causal inference literature on matching methods has been concentrated on dimension reduction techniques \citep[e.g.,][]{rubin1973matching,rubin1973use,rubin1976multivariate,cochran1973controlling}. In this literature, the leading approach for dimension reduction uses the propensity score, which is the conditional probability of treatment given covariate information.
Propensity score methods are designed for calculating average treatment effects (as opposed to conditional average treatment effects) and do not produce exact or almost-exact matches. When treatment is binary, they project data onto one dimension, and closeness of units in propensity score does not imply their closeness in covariate space. As a result, the matches cannot directly be used for estimating heterogeneous treatment effects.

Other causal inference methods have been studied in the literature \citep{gu1993, imbens2004nonparametric}, but almost all of them suffer from at least one of four possible problems: using a black box model that is uninterpretable (i.e., almost all machine learning methods), having a distance metric that is predefined (rather than learned), computational inefficiency, or not being applicable to CATE estimation (as we discussed with propensity scores). 
These issues cause the vast majority of matching methods to be ineffective in producing high quality interpretable CATE estimates. Regression methods can be used for CATE estimation, but only when the regression method is correctly specified -- or in the case of doubly robust estimation \citep[e.g.,][]{farrell2015robust}, either the propensity model or the outcome model needs to be correctly specified. Machine learning approaches generalize regression approaches and can create models that are extremely flexible and predict outcomes accurately for both treatment and control groups \citep{hill2011bayesian,10.1111/ectj.12097,hahn2017bayesian}. However, complicated regression methods lose the interpretability inherent to almost-exact matches and are difficult to troubleshoot and trust.
In practice, MALTS performs similarly to (or better than) several machine learning methods in our experiments, despite being restricted to interpretable almost-exact matches with an interpretable distance metric.

A flexible setup for producing high-quality matches is provided by the optimal matching literature \citep{rosenbaum2016imposing}. These are built on network flow algorithms and integer programming to produce matches that are constrained in user-defined ways \citep{zubizarreta2012using,zubizarreta2014matching,keele2014optimal,resa2016evaluation,pmlr-v54-kallus17a,MorucciEtAl2022}. In all of these approaches, the user defines the distance metric (rather than learning it from data), potentially leading to poor quality matched groups. 
An alternative to optimal matching is coarsened exact matching \citep[CEM, ][]{iacus2012causal}, an approach that requires users to specify explicit bins for all covariates on which to construct matches. 
This requires users to know in advance that the outcomes are insensitive to movements within many high-dimensional bins, 
which is essentially equivalent to the user knowing the answer to the problem we investigate in this work. Large amounts of user choice to define these bins can also lead to unintentional user bias. By \textit{learning} the stretching rather than asking the user to define it as in CEM, this bias is potentially reduced. 

\cite{zhao2004} and \cite{imbens2004nonparametric} discuss the choice of distance metric for matching. The approach by \cite{zhao2004} depends on the correlations between treatment choice, outcome and covariates. However, this approach assumes a model for the relationship between the outcome and covariates, or the treatment choice and covariates. Hence, under model misspecification, the estimator may not be consistent. MALTS learns a distance metric without any model assumptions. 

The present work builds on work of  \cite{wang2017flame, DiengEtAl2018} where a discrete distance metric is learned by considering the prediction quality of the covariate sets. That work does not pertain to continuous covariates, whereas ours does.
There is substantial work on learning distance metrics \citep[though not for causal inference, e.g.,][]{goldberger2005neighbourhood,weinberger2006distance,weinberger2009distance}, where the goal is to learn a distance metric in latent space to separate different classes of data in supervised learning, often with a margin. This is different from our goal of matching for causal inference, but some of our proofs were inspired by this work in supervised learning. 

A sister work, developed in parallel, is that of \citet{AHB}, which learns adaptively-sized hyperboxes as matched groups. MALTS was previously used on the ACIC 2018 Causal Inference Challenge Data \citep[see][]{harsh2019acic}. An extension of MALTS for multi-level treatments has been used to study the effect of seizures on the discharge status of critical ill patients \citep[see][]{parikh2022interpretable}. 	\section{Learning-to-Match Framework}\label{sec:framework}
Within this framework, we perform treatment effect estimation using following three stages: 1) learning a distance metric, 2) matching samples, and 3) estimating CATEs. 

We denote the $p$ dimensional covariate vector space as $\mathcal{X}\subset \mathbb{R}^p$ and the unidimensional outcome space by $\mathcal{Y} \subset \mathbb{R}$. Let $\mathcal{T}$ be a finite label set of treatment indicators (in this paper we consider only the binary case). Let $\mathcal{Z}=\mathcal{X}\times\mathcal{Y}\times\mathcal{T}$ such that $z=(\x,y,t)\in \mathcal{Z}$ means that $\x\in\mathcal{X}$, $y\in\mathcal{Y}$ and $t\in\mathcal{T}$. Let $\mu$ be an unknown probability distribution over $\mathcal{Z}$ such that $\forall z \in \mathcal{Z}, ~ \mu(z)>0$. We assume that $\mathcal{X}$ is a compact convex space with respect to $\|\cdot\|_2$, thus there exists a constant $\mathbf{C}_x$ such that $\|\x\|_2\leq\mathbf{C}_x$. Also, $|y|\leq\mathbf{C}_y$. A distance metric is a symmetric, positive definite function with two arguments from $\mathcal{X}$ such that $\dis: \mathcal{X}\times\mathcal{X} \to \mathbb{R}^+$. A distance metric must obey the triangle inequality. Let $\mathcal{S}_n$ denote a set of $n$ observed units $\{s_1,...,s_n\}$ drawn i.i.d$.$ from $\mu$ such that $\forall i, ~s_i \in \mathcal{Z}$. We parameterize $\dis$ with parameter $\mathcal{M}(\cdot)$, explicitly calling it $\dis_\mathcal{M}$, and let $\mathcal{M}(\mathcal{S}_n)$ denote the parameter learned using $\textsc{MALTS}$ methodology which is described in Section~\ref{sec:method}. 
For ease of notation, we will denote the observed sample of treated units as $\mathcal{S}^{(T)}_n := \{s^{(T)}_i = (\x_i,y_i,t_i)~|~ s^{(T)}_i \in \mathcal{S}_n \text{ and } t_i = T \}$ and the observed sample of control units as $\mathcal{S}^{(C)}_n := \{s^{(C)}_i = (\x_i,y_i,t_i)~|~ s^{(C)}_i \in \mathcal{S}_n \text{ and } t_i = C \}$.

We assume no unobserved confounders and standard ignorability assumptions, i.e.,  $\forall i,~ (Y_i^{(T)},Y_i^{(C)})~ \indep ~T_i ~|~ (X_i=\x_i)$ \citep{Rubin2005} where $Y_i^{(T)}$ and $Y_i^{(C)}$ are potential outcomes for unit $i$ under treatments $(T)$ and $(C)$ respectively, $T_i$ is unit $i$'s treatment choice and $X_i$ corresponds to the vector of covariates for unit $i$. For each individual unit $s_i = (\x_i,y_i,t_i) \in \mathcal{Z}$ we define its conditional average treatment effect (or individualized treatment effect) as the difference of potential outcomes of unit $i$ under the treatment and control, $\tau(\x_i) =$ $\mathbbm{E}\left[Y_i^{(T)} - Y_i^{(C)} | X_i=\x_i \right]$ $=\mathbbm{E}\left[Y_i^{(T)}|X_i=\x_i \right] - \mathbbm{E}\left[Y_i^{(C)} | X_i=\x_i \right]$. 
We use the $\widehat{Y}^{(t)}_{\x_i}$ to refer to the estimated conditional average potential outcome, $\mathbbm{E}\left[Y_i^{(t)} | X_i=\x_i \right]$, for treatment $t\in\mathcal{T}$ and covariate level $\x_i\in\mathcal{X}$. $\widehat{\tau}(\x_i)$ refers to the estimated conditional average treatment effect for covariate value $\x_i$.

Our goal is to minimize the expected loss between estimated treatment effects $\widehat{\tau}(\x)$ and true treatment effects $\tau(\x)$ across target population $\mu(z)$ (this can either be a finite or super-population).

Let the population expected loss be:
\begin{eqnarray*}
\E\left[ \ell(\widehat{\tau}(\x),\tau(\x))\right]
=
\int \ell(\widehat{\tau}(\x),\tau(\x))d\mu
=
\int \ell(\hat{Y}^{(T)}_\x-\hat{Y}^{(C)}_\x, \mathbb{E}[{Y}^{(T)}-{Y}^{(C)}|X=\x]) d\mu.
\end{eqnarray*}

For a finite random i.i.d$.$ sample $\{s_i=(\x_i,y_i,t_i)\}^n_{i=1}$ from the distribution $\mu$, the finite sample version of the average loss can be written as
\begin{eqnarray*}
\frac{1}{n}\sum_{i=1}^n \ell\left(\hat{Y}^{(T)}_{\x_i}-\hat{Y}^{(C)}_{\x_i}, \mathbb{E}\left[ Y_i^{(T)} | X_i = \x_i \right] - \mathbb{E}\left[ Y_i^{(C)} | X_i = \x_i \right]\right).
\end{eqnarray*}
However, we do not observed true values of $\mathbb{E}\left[ Y_i^{(T)} | X_i = \x_i \right]$ and $\mathbb{E}\left[ Y_i^{(C)} | X_i = \x_i \right]$. 

Instead, we could estimate the upper bound of sample average loss as \begin{eqnarray*}
\frac{1}{n}\sum_{i=1}^n t_i \ell\left(\hat{Y}^{(T)}_{\x_i}, y_i\right)+ (1-t_i)\ell\left(\hat{Y}^{(C)}_{\x_i}, y_i\right).
\end{eqnarray*}
Here, we use can $y_i$ for $t_i=1$ as the unbiased estimate of $\mathbb{E}\left[ Y_i^{(T)} | X_i = \x_i \right]$ and similar for $t_i=0$.

 For a unit $s_i$, we estimate the conditional average potential outcomes, $\hat{Y}^{(T)}_{\x_i}$ and $\hat{Y}^{(C)}_{\x_i}$, using the treated and control units' outcomes in the constructed \textit{matched group} using the observed data. 
 The \textit{matched group} $\MG$ of unit $s_i$ for treatment $t'$ under the distance metric $\dis_{\mathcal{M}}$ on covariate space is defined as a set of $K$ nearest neighbors of $s_i$ from set $\mathcal{S}^{(t')}_n = \{ s_k | t_k = t', s_k\in\mathcal{S}_n \}$.
\begin{eqnarray}\label{eqn:mg}
\MG(s_i,\dis_{\mathcal{M}},\mathcal{S}^{(t')}_n,K) =
KNN^{\mathcal{S}_n}_{\mathcal{M}}(\x_i,t') := \bigg\{s_k:\bigg[\sum_{s_l\in\mathcal{S}^{(t')}_n}\mathbbm{1}\Big(\dis_{\mathcal{M}}(
\x_l,\x_i)<\dis_{\mathcal{M}}(\x_k,\x_i)\Big)\bigg] < K \bigg\}.
\end{eqnarray}
We allow reuse of units in multiple matched groups. Thus for a chosen estimator $\phi$,
\begin{equation}\label{eqn:yhat}
\hat{Y}^{(t')}_{\x_i} = \phi\left(\MG(s_i,\dis_{\mathcal{M}},\mathcal{S}^{(t')}_n,K)\right)
\end{equation}
where $K$ is the size of the matched group $\MG(s_i,\dis_{\mathcal{M}},\mathcal{S}^{(t')}_n,K)$. A simple example of $\phi$ is the mean estimator, i.e. $\phi\left(\MG(s_i,\dis_{\mathcal{M}},\mathcal{S}^{(t')}_n,K)\right) =  \frac{1}{K}\sum_{k\in\MG(s_i,\dis_{\mathcal{M}},\mathcal{S}^{(t')}_n,K)} y_k$. However, one can choose the estimator to be a weighted mean, linear regression or a non-parametric model like random-forest, within the matched group. 

Our framework performs honest causal inference by learning a distance metric
from a separate training set of data (not the estimation data considered in the averages above), and
we denote this training set by $\mathcal{S}_{tr}$. To learn $\dis_{\mathcal{M}}$, we minimize the following:
\begin{eqnarray*}
\mathcal{M}(\mathcal{S}_{tr})\in
\textrm{arg}\min_{\mathcal{M}}\left[ 
\begin{array}{l}
\sum_{s_i\in \mathcal{S}^{(T)}_{tr}}
\left| y_i - \hat{Y}^{(T)}_{\x_i}
\right|
+ \sum_{s_i\in \mathcal{S}^{(C)}_{tr}}
\left| y_i - \hat{Y}^{(C)}_{\x_i}
\right|\end{array}
\right],  
\end{eqnarray*}
where $\hat{Y}^{(C)}_{\x_i}$ and $\hat{Y}^{(T)}_{\x_i}$ are defined by Equations (\ref{eqn:mg}) and (\ref{eqn:yhat}) including its dependence on the $\distance$ $\dis_{\mathcal{M}}$, which is parameterized by
 $\mathcal{M}$,
using the training data to create matched groups.

Once $\mathcal{M}(\mathcal{S}_{tr})$ is learned from the training set, it is used for matching (and estimation) on the estimation data.

\subsection{Smooth Distance Metric and Treatment Effect Estimation}
In this subsection, we discuss that if a distance metric is a smooth distance metric, then we can estimate the individualized treatment effect using a finite sample with high probability. First, let us define a smooth distance metric.
\begin{define}
\textbf{(Smooth Distance Metric)}\label{def:smoothdis} $\dis_\mathcal{M}:\mathcal{X}\times\mathcal{X}\to\mathbb{R}^{+}$ 
is a smooth distance metric if there exists a monotonically increasing bounded function 
$\delta_{\dis_\mathcal{M}}(\cdot)$ with zero intercept, 
such that $\forall z_i,z_j \in \mathcal{Z}$ if $t_i=t_j$ and $\dis_\mathcal{M}(\x_i,\x_j)\leq a$ 
then
\begin{equation*}
    \left|\mathbb{E}[Y_i|X_i=\x_i,T_i=t_i]-\mathbb{E}[Y_j|X_j=\x_j,T_j=t_j]\right|\leq\delta_{\dis_\mathcal{M}}(a).
\end{equation*}
\end{define}

The concept of the smooth distance metric is analogous to commonly assumed Lipschitz continuity in the matching literature \citep{abadie2006}.
Note that because the range of $Y$ is bounded, there always exists a choice of the function $\delta_{\mathcal{M}}(\cdot)$ such that a distance metric $d_\mathcal{M}$ is smooth. This choice of $\delta_{\mathcal{M}}(\cdot)$ controls the quality of inference from the matching as we see in Theorem 1 below.

\begin{theorem} \textbf{(Basic CATE Bound for Smooth Distance Metrics)}
\label{th: smoothtau}
Let $\{\mathcal{S}_n\}_{n=1}^\infty$ be a sequence of nested datasets, each of which includes $n$ i.i.d$.$ samples from $\mu(\mathcal{Z})$, $n=1..\infty$. 
Given a smooth distance metric $\dis_{\mathcal{M}}$, covariate vector $\x$, and $\alpha>0$, 
if there exists a small enough value of ``$a$'' and a large enough value of $N$ 
such that $\mathcal{K}^{(t')}_n(\x) = \{z_k : \dis_\mathcal{M}(\mathbf{X}_k,\x)<a, T_k = t', z_k\in \mathcal{S}_n\}$ is non-empty and  $\alpha > 2 \delta_{\dis_\mathcal{M}}(a)$ for all $n\geq N$ and $t'\in\mathcal{T}$,
then

$$ P_{\{Y_i\}_{i=1}^{n} \sim \mu(\mathcal{Y}^n)}\left( |\hat{\tau}(\x) - \tau(\x)| \geq \alpha
\right) 
\leq 4\exp\left(\frac{- K_n(\x)(\frac{\alpha}{2}-\delta_{\dis_\mathcal{M}}(a))^2}{2\mathbf{C}_y}\right) $$ where $\hat{\tau}(\x)$ is the estimated conditional average treatment effect using the matched sets $\mathcal{K}^{(1)}_n(\x)$ and $\mathcal{K}^{(0)}_n(\x)$, $\tau(\x)$ is the true conditional average treatment effect, $K_n(\x) = \min_{t'} |\mathcal{K}^{(t')}_n(\x)|$, and $\delta_{\dis_\mathcal{M}}(a)$ is the bound from Definition~\ref{def:smoothdis} (definition of smooth distance metric).
\end{theorem}

Theorem~\ref{th: smoothtau} directly follows from Lemma~\ref{lm: smoothy} in the \ref{sec:appendixA} which proves that for all $t' \in \mathcal{T}$ and $\x\in\mathcal{X}$, we can estimate average conditional potential outcomes, $\mathbb{E}[Y^{(t')}|X=\x]$, correctly with high probability using nearest neighbor matching under any smooth distance metric, and Lemma~\ref{lm: ytotau} in \ref{sec:appendixA} which proves that estimating average conditional potential outcomes correctly with high probability leads to estimating CATEs, $\tau$, correctly with high probability.
 
 Our setup and Definition~\ref{def:smoothdis} are similar to one described by \cite{kara2017data}. Our result in Lemma~\ref{lm: smoothy} proves the consistency for a uniform weighted nearest neighbor estimator where the weights are probability weights. The result is in congruence with consistency results by \cite{stone1977consistent} and \cite{jiang2019non}; those works handled the special case where the weights are uniform probability weights instead of \textit{any} probability weights.

  Note that matching using any type of stretch norm that induces a smooth distance metric, including Mahalanobis distance (or its special case with an identity covariance matrix, the $L_2$ distance), to adjust for confounding produces consistent estimates of average treatment effects. Prognostic score \citep{hansen2008prognostic} and other approaches that induce a smooth distance metric also produce consistent estimates of ATE.

\section{Matching After Learning to Stretch (MALTS)}\label{sec:method}
\normalsize
MALTS performs weighted nearest neighbors matching, where the weights for the nearest neighbors can be learned by minimizing the following objective. This objective is simply the loss of the in-sample nearest neighbor estimator:

\footnotesize
\begin{eqnarray}\label{eq: obj}
   \mathbf{W} \in \textrm{arg}\min_{\widetilde{\mathbf{W}}} \left[ \sum_{i \in \mathcal{S}^{(T)}_{tr}} \left\|y_{i} - \sum_{s_l \in \mathcal{S}^{(T)}_{tr}, i\neq l} \widetilde{W}_{i,l} y_{l}\right\| \right]
    &+& \left[\sum_{i \in \mathcal{S}^{(C)}_{tr}}  \left\|y_{i} - \sum_{l \in \mathcal{S}^{(C)}_{tr}, i\neq l} \widetilde{W}_{i,l} y_{l}\right\| \right]+ \textrm{Reg}(\widetilde{W}),
\end{eqnarray}
\normalsize
where $\textrm{Reg}(\cdot)$ is a regularization function. We let $\widetilde{W}_{i,l}$ be a function of $\dis_{\mathcal{M}}(\x_i,\x_l)$. For example, the $\widetilde{W}_{i,l}$ can encode whether $l$ belongs to $i$'s $K$-nearest neighbors. Alternatively, they can encode soft $\textrm{KNN}$ weights where $\widetilde{W}_{i,l} \propto e^{-\dis_{\mathcal{M}}(\x_i,\x_l)}$. Thus, the intuition is to learn $\mathbf{W}$ such that the in-sample nearest-neighbors estimator is as accurate as possible.

As a reminder of our notation, we consider distance metric $\dis_{\mathcal{M}}$ parameterized by a set of parameters $\mathcal{M}$.
We use Euclidean distances for continuous covariates, namely distances of the form $\|\mathcal{M} \mathbf{x}_a -\mathcal{M} \mathbf{x}_b\|_2$ where $\mathcal{M}$ encodes the orientation of the data. In the past, $\mathcal{M}$ has been hard-coded rather than learned; an example in the causal inference literature is the classical Mahalanobis distance ($\mathcal{M}$ is fixed as the inverse covariance matrix for the observed covariates). This approach has been demonstrated to perform well in settings where all covariates are observed and the inferential target is the average treatment effect \citep{stuart2010matching}. We are interested instead in individualized treatment effects, and just as the choice of Euclidean norm in Mahalanobis distance matching depends on the estimand of interest, the stretch metric needs to be amended for this new estimand. We propose learning the parameters of a distance metric, $\mathcal{M}$, directly from the observed data rather than setting it beforehand.
The parameters of distance metric $\mathcal{M}$ can be learned such that $\mathbf{W}$ minimizes the objective function on the training set.

In our framework, we can define ``approximate closeness" differently for discrete covariates if desired. For continuous covariates, MALTS uses Euclidean distance, which is also a reasonable metric to use for binary data  \citep[e.g., Mahalanobis-distance-matching papers recommend converting unordered categorical variables to binary indicators, see][]{stuart2010matching}; however, there are benefits to using other metrics, such as weighted Hamming distances, for comparison among sets of binary covariates. 
To accommodate a combination of Euclidean and Hamming distances,
we parameterize our distance metric in terms of two components: one is a learned weighted Euclidean distance for continuous covariates while the other is a learned weighted Hamming distance for discrete covariates as in the FLAME and DAME algorithms \citep{wang2017flame, DiengEtAl2018}. These components are separately parameterized by matrices $\mathcal{M}_c$ and $\mathcal{M}_d$ respectively, $\mathcal{M} = \left[\mathcal{M}_c, \mathcal{M}_d\right]$ (here $c$ indicates ``continuous,'' and $d$ indicates ``discrete''). Let $a = (a_c,a_d)$ and $b = (b_c,b_d)$ be the covariates for two individuals split into continuous and discrete pairs respectively. 

\noindent\textbf{Operationalizing Equation~\eqref{eq: obj}:} To perform the step called ``Distance Metric Learning'' in Figure~\ref{fig:malts_framework} we propose the following form for the distance metric:
$$\dis_\mathcal{M}(a,b) = d_{\mathcal{M}_c}(a_c,b_c) + d_{\mathcal{M}_d}(a_d,b_d) \text{, where}$$
  $$d_{\mathcal{M}_c}(a_c,b_c) = \|\mathcal{M}_{c}a_c - \mathcal{M}_{c}b_c\|_{2}, \hspace{0.25 cm}d_{\mathcal{M}_d}(a_d,b_d) = \sum_{j=0}^{|a_d|} \mathcal{M}_d^{(j,j)} \mathbbm{1}[a_d^{(j)}\neq b_d^{(j)}], $$
and $\mathbbm{1}[A]$ is the indicator that event $A$ occurred. 
We thus perform learned Hamming distance matching on the discrete covariates and learned-Mahalanobis-distance matching for continuous covariates. 

MALTS performs an ``honest'' causal inference by splitting the observed sample dataset $\mathcal{S}_n$ into a training set $\mathcal{S}_{tr}$ (not for matching) and an estimation set $\mathcal{S}_{est}$ (for matching).
We learn $\mathcal{M}(\mathcal{S}_{tr})$ using the training set $\mathcal{S}_{tr}$ such that in Equation~\eqref{eq: obj}, $\widetilde{W}_{i,l} = \frac{e^{-\dis_\mathcal{M}(\x_i,\x_l)}}{\sum_{s_k \in \mathcal{S}^{(t_i)}_{tr}} e^{-\dis_\mathcal{M}(\x_i,\x_k)}}$ and $Reg(\widetilde{W})=\|\mathcal{M}\|_{\mathcal{F}}$ which defines \textbf{MALTS' main implemented optimization problem}:
\begin{equation}
    \mathcal{M}(\mathcal{S}_{tr}) \in \argmin_{\mathcal{M}} \left( c\|\mathcal{M}\|_\mathcal{F} + \Delta^{(C)}_{\mathcal{S}_{tr}}(\mathcal{M}) + \Delta^{(T)}_{\mathcal{S}_{tr}}(\mathcal{M}) \right) 
\end{equation}
where
$\|\cdot\|_{\mathcal{F}}$ is the Frobenius norm of the matrix, and:
\begin{equation}
\label{eq:delta_obj}
\begin{split}
    \Delta^{(t)}_{\mathcal{S}_{tr}}(\mathcal{M}) :&= \frac{1}{|\mathcal{S}^{(t)}_{tr}|} \sum_{s_i\in\mathcal{S}^{(t)}_{tr}}\left| y_i - \sum_{s_l \in \mathcal{S}^{(t)}_{tr}} \frac{e^{-\dis_\mathcal{M}(\x_i,\x_l)}}{\sum_{s_k \in \mathcal{S}^{(t)}_{tr}} e^{-\dis_\mathcal{M}(\x_i,\x_k)}}y_l  \right| \\ &= \frac{1}{|\mathcal{S}^{(t)}_{tr}|}\sum_{s_i\in\mathcal{S}^{(t)}_{tr}}\left|\sum_{s_l \in \mathcal{S}^{(t)}_{tr}} \frac{e^{-\dis_\mathcal{M}(\x_i,\x_l)}}{\sum_{s_k \in \mathcal{S}^{(t)}_{tr}} e^{-\dis_\mathcal{M}(\x_i,\x_k)}}(y_i-y_l)  \right|.
\end{split}
\end{equation}

\noindent\textbf{Matching and Estimation:} To perform the step called ``Nearest Neighbor Matching,'' which produces ``Matched Groups'' that are used to estimate ``CATEs'' in Figure~\ref{fig:malts_framework}, we use the learned distance metric $\mathcal{M}(\mathcal{S}_{tr})$. To estimate conditional average treatment effects (CATEs) for each unit in the estimation set, we use its nearest neighbors from the same estimation set. Specifically, for any given unit $s$ in the estimation set, we construct a K-nearest neighbor matched group $\MG(s,\dis_{\mathcal{M}(\mathcal{S}_{tr})},\mathcal{S}_{est},K)$ using control set $\mathcal{S}^{(C)}_{est}$ and treatment set $\mathcal{S}^{(T)}_{est}$. For a choice of estimator $\phi$, the estimated CATE for a treated unit $s = (\x_s,y_s,t_s=t')$ is calculated as follows: 
$$\hat{\tau}(\x) = \phi
\left(\MG(s,\dis_{\mathcal{M}(\mathcal{S}_{tr})},\mathcal{S}^{(T)}_{est},K)\right) - \phi
\left(\MG(s,\dis_{\mathcal{M}(\mathcal{S}_{tr})},\mathcal{S}^{(C)}_{est},K)\right).$$

A simple example of $\phi$ is the empirical mean, i.e., $$\phi\left(\MG(s,\dis_{\mathcal{M}},\mathcal{S}^{(t)}_n,K)\right)=\frac{1}{K}\sum_{k\in\MG(s,\dis_{\mathcal{M}},\mathcal{S}^{(t)}_n,K)}y_k.$$ However, one can choose the estimator to be a weighted mean, linear regression or a non-parametric model like Random Forest. Particular choices of $\phi$ can also play a role in bias-adjustment to improve the matching estimator of the ATE as in \cite{abadie2011bias}.

For $\phi\left(\MG(s,\dis_{\mathcal{M}},\mathcal{S}_n,K)\right) = \sum_{k\in\MG(s,\dis_{\mathcal{M}},\mathcal{S}_n,K)} \widetilde{W}_{k}y_k$, if $\widetilde{W}_{k}$ is chosen to be proportional to $e^{\dis_{\mathcal{M}}(\x,\x_k)}$, then it leads to multi-robust (defined shortly) and generalizable CATE estimates via soft KNN (as shown in Theorem \ref{th: robust} and Theorem \ref{th: gen} below), while letting $\widetilde{W}_k$ be proportional to $\mathbbm{1}\left[s_k\in\textrm{KNN}^{\mathcal{S}^{(C)}_{est}}_{\mathcal{M}(\mathcal{S}_{tr})}\right]$ produces interpretable matched groups. 

\noindent\textbf{Hyperparameter choice:} MALTS has four main hyperparameters: 1) K, which is the number of nearest neighbors used to estimate the counterfactual, which can be chosen by cross-validation. 2) $n$, the size of training set, i.e., the size of the split on the left of Figure \ref{fig:malts_framework}. This can be chosen based on the amount of data relative to the number of features, though typically we choose it to be 10\% of the data. 3) The maximum allowed diameter or caliper to prune bad matched groups. If the matches have a larger diameter, the matches are not tight and we may not be able to trust their estimates. The maximum diameter can be chosen by domain knowledge; the user defines how far apart points can be to make the matched group less interpretable. 4) The number of repeats refers to the number of times we shuffle the data and re-partition it for MALTS’ training and estimation procedure. A larger number of repeats of the whole process helps with smoothing out the estimates over different train/test splits.  	\section{Robustness and Generalization of MALTS}
\label{sec:theory}
In this section, we show that the MALTS framework correctly estimates the distance metric, facilitating correct estimates of CATEs under SUTVA and a positivity assumption. After basic definitions, and after showing that the learned distance metric and objective are bounded, we introduce and define the concepts of multi-robustness and generalizability of the learned distance metric. Multi-robustness implies that for any possible pair of points the empirical average loss is not far away from the population average loss. Theorem~\ref{th: robust} proves that the distance metric learned by the MALTS algorithm is multi-robust. We use these results along with the error bound shown in Lemma~\ref{lm: whpavgloss}, to show that MALTS' distance metric is generalizable, i.e., the population average loss and the empirical average loss on the observed data for the learned distance metric are close with high probability. Lastly, we show that MALTS' distance metric is asymptotically generalizable and that the empirical average loss approaches the population average loss as the size of the dataset goes to infinity.

\noindent\textbf{Basic definitions of empirical loss and population loss.} 
First, we define a pairwise loss for $s_i$ and $s_l$ so that it is only finite for treatment-treatment or control-control matched pairs,
\begin{equation*}
   loss[\mathcal{M},s_i,s_l] := \begin{cases} 
      e^{-\dis_\mathcal{M}(\x_i,\x_l)}|y_i-y_l| & \text{ if }t_i=t_l \\
      \infty & \text{otherwise}. \\
   \end{cases}
\end{equation*}
This loss is high for pairs of points that are close (i.e., with small $\dis_\mathcal{M}(\x_i,\x_l)$) when the outcomes $y_i$ and $y_l$ values are very different.
Further, we define an empirical average pairwise loss over finite sample $\mathcal{S}_n$ of size $n$ as 
\begin{equation*}
   L_{emp}(\mathcal{M},\mathcal{S}_n) := \frac{1}{n^2}\sum_{(s_i,s_l)\in(\mathcal{S}_n\times\mathcal{S}_n)} loss[\mathcal{M},s_i,s_l]
\end{equation*}
and define an average loss over population $\mathcal{Z}$ as 
\begin{equation*}
   L_{pop}(\mathcal{M},\mathcal{Z}) := \mathbb{E}_{z_i,z_l\overset{i.i.d}{\sim}\mu(\mathcal{Z})} \Big[ loss[\mathcal{M},z_i,z_l] \Big].
\end{equation*}

\noindent\textbf{The search space over distance metrics is bounded.}
We show a basic result about the optimization-based approach we take to learn the distance metric. Specifically, we show that the learned distance metric will be in a bounded region of search space. 

Now, because the learned $\mathcal{M}(\mathcal{S}_{tr})$ on the set $\mathcal{S}_{tr}$ is the distance metric that minimizes the given objective function, we know that the following inequality is true, which states that the learned parameter has a lower training objective than that of the trivial parameter $\mathbf{0}$:
\begin{equation}
    \Big( c\|\mathcal{M}(\mathcal{S}_{tr})\|_\mathcal{F} + \Delta^{(C)}_{\mathcal{S}_{tr}}(\mathcal{M}(\mathcal{S}_{tr})) + \Delta^{(T)}_{\mathcal{S}_{tr}}(\mathcal{M}(\mathcal{S}_{tr})) \Big) \leq  \Big( c\|\mathbf{0}\|_\mathcal{F} + \Delta^{(C)}_{\mathcal{S}_{tr}}(\mathbf{0}) + \Delta^{(T)}_{\mathcal{S}_{tr}}(\mathbf{0}) \Big) =: g_0.
\end{equation}

Denoting the right hand side of the inequality by $g_0$ we note that we can limit our search space over distance metrics $\mathcal{M}$ that satisfy the following inequality:
\begin{equation*}
    \|\mathcal{M}\|_\mathcal{F} \leq \frac{g_0}{c}.
\end{equation*}

\noindent\textbf{The objective function terms are bounded.} The objective terms $\Delta_{\mathcal{S}_{tr}}^{(C)}$ and $\Delta_{\mathcal{S}_{tr}}^{(T)}$ (defined in Equation~\eqref{eq:delta_obj}) for learning the distance metric are also bounded, although it is not that easy to see this directly because their denominators are somewhat complicated, involving a sum over exponential terms. Here, we point out that because the learned distance metric is bounded, the objective's terms ($\Delta_{\mathcal{S}_{tr}}^{(C)}$ and $\Delta_{\mathcal{S}_{tr}}^{(T)}$) are also bounded. Specifically, their upper bound is proportional to the empirical average pairwise losses $L_{emp}(\mathcal{M},\mathcal{S}^{(C)}_{tr})$ and $L_{emp}(\mathcal{M},\mathcal{S}^{(T)}_{tr})$, defined above. Further, in Theorem~\ref{th: gen}, we show that for $t'\in\{T,C\}$ the empirical average loss $L_{emp}(\mathcal{M},\mathcal{S}^{(t')}_{tr})$ is close to population average pairwise loss $L_{pop}(\mathcal{M}(\mathcal{S}_n),\mathcal{Z}^{(t')})$ with high probability. Following  Equations~\eqref{eq:delta_upperbound_c} and \eqref{eq:delta_upperbound_t} and Theorem~\ref{th: gen}, the objective terms $\Delta_{\mathcal{S}_{tr}}^{(C)}(\mathcal{M})$ and $\Delta_{\mathcal{S}_{tr}}^{(T)}(\mathcal{M})$ are upper-bounded by a term proportional to the population average pairwise loss with high probability.
\begin{eqnarray*}
    \Delta^{(C)}_{\mathcal{S}_{tr}}(\mathcal{M}) &\leq& \frac{1}{|\mathcal{S}^{(C)}_{tr}|}\sum_{s_i\in\mathcal{S}^{(C)}_{tr}}\sum_{s_l \in \mathcal{S}^{(C)}_{tr}}\left| \frac{e^{-\dis_\mathcal{M}(\x_i,\x_l)}}{\sum_{s_k \in \mathcal{S}^{(C)}_{tr}} e^{-\dis_\mathcal{M}(\x_i,\x_k)}}(y_i-y_l)  \right|\\ &=& \;\; \frac{1}{|\mathcal{S}^{(C)}_{tr}|}\sum_{s_i\in\mathcal{S}^{(C)}_{tr}} \frac{\sum_{s_l \in \mathcal{S}^{(C)}_{tr}}loss[\mathcal{M},s_i,s_l]}{\sum_{s_k \in \mathcal{S}^{(C)}_{tr}} e^{-\dis_\mathcal{M}(\x_i,\x_k)}}.
\end{eqnarray*}
We know that:
\begin{equation*}
    \forall i,k ~ \dis_\mathcal{M}(\x_i,\x_k) = \left[(\x_i - \x_k)'\mathcal{M}'\mathcal{M}(\x_i - \x_k)\right]^{1/2} \leq \|\x_i - \x_k\|_{2} \|\mathcal{M}\|_{\mathcal{F}} \leq \frac{g_0\mathbf{C}_x^2}{c}.
\end{equation*}
Together, the two previous lines imply:
\begin{equation}\label{eq:delta_upperbound_c}
\Delta^{(C)}_{\mathcal{S}_{tr}}(\mathcal{M}) 
\leq \frac{1}{\exp{(-\frac{g_0\mathbf{C}_x^2}{c})}\left|\mathcal{S}^{(C)}_{tr}\right|^2}\sum_{s_i\in\mathcal{S}^{(C)}_{tr}} \sum_{s_l \in \mathcal{S}^{(C)}_{tr}} loss[\mathcal{M},s_i,s_l] = \frac{ L_{emp}(\mathcal{M},\mathcal{S}^{(C)}_{tr})}{\exp{(-\frac{g_0\mathbf{C}_x^2}{c})}}.
\end{equation}
Similarly for the treatment units, we have 
\begin{equation}\label{eq:delta_upperbound_t}
\Delta^{(T)}_{\mathcal{S}_{tr}}(\mathcal{M}) \leq \frac{ L_{emp}(\mathcal{M},\mathcal{S}^{(T)}_{tr})}{\exp{(-\frac{g_0\mathbf{C}_x^2}{c})}}.
\end{equation}

Now, we define a few concepts important for our results including covering number, multi-robustness, and generalizability. The following definitions and results closely align with the theoretical guarantees of distance metric learning algorithms in \cite{bellet2015robustness} and \cite{xu2012robustness}. Our work extends these results to learn a  distance metric for causal inference.

\begin{define}
(\textbf{Covering Number})
Let ($\mathcal{U},\dis$) be a metric space. Consider a subset $\mathcal{V}$ of $\mathcal{U}$, then $\hat{\mathcal{V}} \subset \mathcal{V}$ is called a $\gamma$-cover of $\mathcal{V}$ if for any $v \in \mathcal{V}$, we can always find a $\hat{v}\in\hat{\mathcal{V}}$ such that $\dis(v,\hat{v})\leq\gamma$. Further, the $\gamma$-covering-number of $\mathcal{V}$ under the distance metric $\dis$ is defined by
$\mathbf{N}(\gamma,\mathcal{V},\dis) := \min\big\{ |\hat{\mathcal{V}}| ~:~ \hat{\mathcal{V}} \text{ is a }\gamma\text{-cover of }\mathcal{V} \big\}$.
\end{define}
Note that $\mathbf{N}(\gamma,\mathcal{V},\dis)$ is finite if $\mathcal{U}$ is compact.

\begin{define}
(\textbf{Robustness})
\label{def:robust}
A learned distance metric $\mathcal{M}(\cdot)$ is $(K,\epsilon(\cdot))$-robust for a given $K$ and $\epsilon(\cdot):(\mathcal{Z}\times\mathcal{Z})^n \to \mathbb{R}$, if we can partition $\mathcal{X}$ into $K$ disjoint sets $\{C_i\}_{i=1}^K$ such that for any subsample $\mathcal{S}_{tr}$ and its corresponding pair set $\mathcal{S}_{tr}^2 := \mathcal{S}_{tr} \times \mathcal{S}_{tr}$, we have for any pair of training units
$\big(s_1=(\x_1,y_1,t_1),s_2=(\x_2,y_2,t_2)\big)\in\mathcal{S}_{tr}^2$, and for any pair of units in the support $\big(z_1=(\x'_1,y'_1,t'_1),z_2=(\x'_2,y'_2,t'_2)\big)\in\mathcal{Z}^2,~\forall i,l \in \{1,...,K\}$, 
$$\text{if } \x_1,\x'_1 \in C_i \text{ and }  \x_2,\x'_2 \in C_l \text{ such that } t_1=t'_1=t_2=t'_2  \text{ then }$$ $$\Big|~ loss[\mathcal{M}(\mathcal{S}_{tr}),s_1,s_2] - loss[\mathcal{M}(\mathcal{S}_{tr}),z_1,z_2]~ \Big|\leq \epsilon(\mathcal{S}_{tr}).$$
\end{define}
Intuitively, \textit{robustness} means that for any possible unit in the support, the loss is not far away from the loss of nearby units in the training set, should some training units exist nearby.  \citep[This terminology is aligned with the distance metric learning literature, e.g.,][and it is different from robustness to model misspecification that frequently appears in the causal inference literature in terms such as ``doubly robust estimator.'']{bellet2015robustness,xu2012robustness}

\begin{define}
(\textbf{Multi-Robustness})
\label{def:multirobust}
\\A learned distance metric $\mathcal{M}(\cdot)$ is $(K,\epsilon(\cdot))$-multirobust for a given $K$ and $\epsilon(\cdot):\mathcal{Z}^n \to \mathbb{R}$, if we can partition $\mathcal{X}$ into $K$ disjoint sets $\textbf{C} = \{C_i\}_{i=1}^K$ such that for any subsample $\mathcal{S}_n$ and its corresponding pair set $\mathcal{S}_n^2 := \mathcal{S}_n \times \mathcal{S}_n$, we have
$\forall\big(s_1=(x_1,y_1,t_1),s_2=(x_2,y_2,t_2)\big)\in\mathcal{S}_n^2,~\forall\big(z_1=(x'_1,y'_1,t'_1),z_2=(x'_2,y'_2,t'_2)\big)\in\mathcal{Z}^2,~\forall i,l \in \{1,...,K\}$, 
\begin{eqnarray*}
\left.\begin{aligned}
&
\mathrm{given } \ \  \widehat{\overline{loss}}[\mathcal{M}(\mathcal{S}_n),C^{(t')}_i,C^{(t')}_l] := \frac{1}{|C^{(t')}_i| |C^{(t')}_l|}\sum_{(s_i,s_l)\in C^{(t')}_i\times C^{(t')}_l} loss[\mathcal{M}(\mathcal{S}_n),s_1,s_2] & \\ 
&
\mathrm{ and } \ \  \overline{loss}[\mathcal{M}(\mathcal{S}_n),C^{(t')}_i,C^{(t')}_l] := \mathbbm{E}[loss(\mathcal{M},Z_i,Z_l) ~|~ X_i^\prime\in C_i^{(t^\prime)}, X_l^\prime\in C_l^{(t^\prime)}]\\
& \forall C_i,C_l \in \textbf{C}, \ \ \Big|~ \widehat{\overline{loss}}[\mathcal{M}(\mathcal{S}_n),C^{(t')}_i,C^{(t')}_l] - \overline{loss}[\mathcal{M}(\mathcal{S}_n),C^{(t')}_i,C^{(t')}_l]~ \Big|\leq \epsilon(\mathcal{S}_n). & 
\end{aligned}\right.
\end{eqnarray*}
\end{define} 

Intuitively, \textit{multi-robustness} means that for any possible pair of points from any two partitions of $\mathcal{X}$, the empirical average loss over training points is not far away from the population average loss. As the training procedure aims at minimizing the total loss, we can safely say that a multi-robust method will not perform poorly out of sample.

\begin{define}
(\textbf{Generalizability}) 
\label{def:general}\\
A learned distance metric $\mathcal{M}(\cdot)$ is said to generalize with respect to the given training sample $\mathcal{S}_n$ if 
$$ P_{\mathcal{S}_n}\left(\sum_{t'\in\mathcal{T}}\Big| L_{pop}(\mathcal{M}(\mathcal{S}_n),\mathcal{Z}^{(t')}) - L_{emp}(\mathcal{M}(\mathcal{S}_n),\mathcal{S}^{(t')}_n) \Big| \geq \epsilon \right) \leq \delta_\epsilon$$
where $\delta_\epsilon$ is a decreasing function of $\epsilon$ with zero-intercept.
\end{define}

\begin{define}
(\textbf{Asymptotic Generalizability})\\ 
A learned distance metric $\mathcal{M}(\cdot)$ is said to asymptotically generalize with respect to the given training sample $\mathcal{S}_n$ if 
$$ \lim_{n\to\infty}\sum_{t'\in\mathcal{T}}\Big| L_{pop}(\mathcal{M}(\mathcal{S}_n),\mathcal{Z}^{(t')}) - L_{emp}(\mathcal{M}(\mathcal{S}_n),\mathcal{S}^{(t')}_n) \Big| = 0$$
\end{define}

Given these definitions, we first show that the distance metric learned using \textsc{MALTS} is robust in Theorem~\ref{th: robust} and we extend the argument to show that it is also generalizable in Theorem~\ref{th: gen}. 

\begin{theorem} \textbf{(MALTS' learned distance metric is multi-robust)}
\label{th: robust}
With probability greater than $\left(1 - \exp\left(-\frac{\beta^2 \left(\rho^{(t')}_{\gamma}\right)^2}{ n^{(t')} B^2 }\right)\right)$, the distance metric $\mathcal{M}(\cdot)$ learned using \textsc{MALTS} is ${\Bigg(\mathbf{N}(\gamma,\mathcal{X},\|\cdot\|_2),\beta\Bigg)\mathrm{-multirobust}}$
for arbitrary chosen values of $\gamma>0$ and $\beta\geq0$,
where $B$ is $\max_{z_1,z_2} loss(\mathcal{M}(\mathcal{S}_n),z_1,z_2)$, $\{C_i\}_{i=1}^{K}$ is the partition of $\mathcal{X}$ into non-empty sets $C_i$'s such that $K$ is the $\gamma$-covering number of $\mathcal{X}$, $C_i^{(t')} = \{z_j=(\x_j,y_j,t_j) : t_j=t',\x_j\in C_i \}$ and $\rho^{(t')}_{\gamma} = \min_i |C^{(t')}_i|$.
\end{theorem}
\textbf{Proof (Theorem \ref{th: robust}). } Given $\mathcal{Z}=\mathcal{X}\times\mathcal{Y}\times\mathcal{T}$, we consider the following definition of a minimum sized $\gamma$-cover $\hat{\mathcal{V}}$ of the set $\mathcal{X}$ under the distance metrix $\|\cdot\|_2$: Partition the set into $K$ disjoint subsets $\textbf{C}_\gamma = \{C_i\}_{i=1}^{K}$ such that $K$ is the $\gamma$-covering-number of $\mathcal{X}$ under $\|\cdot\|_2$ (which is exactly equal to $|\hat{\mathcal{V}}|$) where each $C_i$ is contained in the $\gamma$-neighborhood of each $\hat{v}_i\in\hat{\mathcal{V}}$ and each $C_i$ contains at least one control and one treated sample. Note that if $\mathcal{X}$ is a compact convex set, then such a cover and the corresponding packing $\textbf{C}_\gamma$ exists and $K=|\textbf{C}_\gamma|$ is finite. 

For any arbitrary $C_i$ and $C_l$ in $\textbf{C}_\gamma$, consider the empirical average loss for all training units $s_i \in C_i$ and $s_l \in C_l$ with treatment $t'$:
\begin{eqnarray*}
\left.\begin{aligned}
& \widehat{\overline{loss}}\left[\mathcal{M}(\mathcal{S}_n),C^{(t')}_i,C^{(t')}_l\right] = \frac{1}{|C^{(t')}_i\|C^{(t')}_l|}\sum_{(s_i,s_l)\in C^{(t')}_i\times C^{(t')}_l} loss[\mathcal{M}(\mathcal{S}_n),s_i,s_l] & 
\end{aligned}\right.
\end{eqnarray*}
and the expected loss for units $Z_i$ and $Z_l$:
\begin{eqnarray*}
\left.\begin{aligned}
&
\overline{loss}\left[\mathcal{M}(\mathcal{S}_n),C^{(t')}_i,C^{(t')}_l\right] = \mathbbm{E}\left[loss(\mathcal{M},Z_i,Z_l)~|~X_i^\prime\in C_i^{(t^\prime)}, X_l^\prime\in C_l^{(t^\prime)}\right]. &
\end{aligned}\right.
\end{eqnarray*}

   Let $f$ be a function of the set of independent random variables such that  $$f(s_1,\dots,s_{|C^{(t')}_i|},s_{|C^{(t')}_l|+1},\dots,s_{|C^{(t')}_i|+|C^{(t')}_l|}) = \frac{1}{|C^{(t')}_i\|C^{(t')}_l|} \sum_{j=1}^{C^{(t')}_i}\sum_{i=C^{(t')}_i+1}^{C^{(t')}_l} loss[\mathcal{M}(\mathcal{S}_n),s_i,s_l].$$ Thus, $f(s_1,\dots,s_{|C^{(t')}_i|},s_{|C^{(t')}_l|+1},\dots,s_{|C^{(t')}_i|+|C^{(t')}_l|}) = \widehat{\overline{loss}}\left[\mathcal{M}(\mathcal{S}_n),C^{(t')}_i,C^{(t')}_l\right]$. 
   
   Now, let $\rho^{(t')}_{\gamma}$ be the density of the $\gamma$-cover for treatment $t'$, defined as the number of units with treatment $t'$ in the smallest partition set $\rho^{(t')}_{\gamma} = \min_i |C^{(t')}_i|$ and $B = \max_{z_1,z_2} loss(\mathcal{M}(\mathcal{S}_n),z_1,z_2)$. Now, we show that $f(\cdot)$ has bounded difference. Without loss of generality, consider an index $j \leq |C^{(t')}_i|$, then
    \begin{align*}
        |f(s_1,\dots,s_{j},\dots,s_{|C^{(t')}_i|+|C^{(t')}_l|}) - f(s_1,\dots,s'_{j},\dots,s_{|C^{(t')}_i|+|C^{(t')}_l|})| 
        \\= \left| \frac{1}{|C^{(t')}_i\|C^{(t')}_l|} \sum_{i=|C^{(t')}_i|+1}^{|C^{(t')}_i|+|C^{(t')}_l|} loss[\mathcal{M}(\mathcal{S}_n),s_i,s_j] - loss[\mathcal{M}(\mathcal{S}_n),s_i,s'_j] \right| 
        \\ \leq \frac{1}{|C^{(t')}_i\|C^{(t')}_l|} \sum_{i=|C^{(t')}_i|+1}^{|C^{(t')}_i|+|C^{(t')}_l|} \left| loss[\mathcal{M}(\mathcal{S}_n),s_i,s_j] - loss[\mathcal{M}(\mathcal{S}_n),s_i,s'_j] \right|
        \\ \leq \frac{1}{|C^{(t')}_i\|C^{(t')}_l|} \sum_{i=|C^{(t')}_i|+1}^{|C^{(t')}_i|+|C^{(t')}_l|} \left| loss[\mathcal{M}(\mathcal{S}_n),s_i,s_j]\right| + \left| loss[\mathcal{M}(\mathcal{S}_n),s_i,s'_j] \right| 
        \\ \leq  \frac{|C^{(t')}_l|}{|C^{(t')}_i\|C^{(t')}_l|} B = \frac{B}{|C^{(t')}_i|} \leq \frac{B}{\rho^{(t')}_{\gamma}}.
    \end{align*}
    
Similarly, for any $j > |C^{(t')}_i| $, $$|f(s_1,\dots,s_{j},\dots,s_{|C^{(t')}_i|+|C^{(t')}_l|}) - f(s_1,\dots,s'_{j},\dots,s_{|C^{(t')}_i|+|C^{(t')}_l|})| \leq \frac{2B}{\rho^{(t')}_{\gamma}}.$$.

As $f()$ is a function of independent $|C^{(t')}_i| + |C^{(t')}_l|$ random variables, by McDiarmid's inequality:
\begin{eqnarray*}
    &&P\left(
        \left|
            \widehat{\overline{loss}}\left[\mathcal{M}(\mathcal{S}_n),C^{(t')}_i,C^{(t')}_l\right] 
            - \overline{loss}[\mathcal{M}(\mathcal{S}_n),C^{(t')}_i,C^{(t')}_l]
        \right|
    \geq\beta\right) 
    \\ &&\leq \text{exp}\left( - \frac{2\beta^2}{\sum_{i=1}^{|C^{(t')}_i| + |C^{(t')}_l|} \frac{B^2}{(\rho^{(t')}_{\gamma})^2} } \right) 
    = \text{exp}\left( - \frac{2\beta^2 \left(\rho^{(t')}_{\gamma}\right)^2}{(|C^{(t')}_i| + |C^{(t')}_l|) B^2 } \right) \leq \text{exp}\left( - \frac{\beta^2 \left(\rho^{(t')}_{\gamma}\right)^2}{ n^{(t')} B^2 } \right).
\end{eqnarray*}
\ensuremath{\hfill\blacksquare}


We will need the following lemma to prove Theorem~\ref{th: gen}. The lemma provides a bound for a particular treatment assignment, while the theorem sums over all treatment assignments.
\begin{lemma}
\label{lm: whpavgloss}
\textbf{(Error Bound)} 
Given sample $\mathcal{S}_{n}\overset{i.i.d}{\sim}\mu(\mathcal{Z})$ where $n^{(t')}$ is the number of units with $t_i=t'$ in $\mathcal{S}_{n}$, and choosing $B>0$ for which $loss[\cdot,z_i,z_l]\leq B$ $\forall z_i,z_l\in\mathcal{Z}$ (B is finite because $\mathcal{X}$ is compact and $\mathcal{Y}$ is bounded): if a learning algorithm provides a distance metric
$\mathcal{M}(\mathcal{S}_{n})$ that is $(K,\epsilon(\cdot))$-multi-robust with probability $p_{mr}(\epsilon)$, then for any $\mathcal{E}>0$, with probability greater than or equal to $(1-\mathcal{E})(p_{mr}(\epsilon))^{K^2}$ we have
$$\forall t'\in\mathcal{T},~\Big| L_{pop}(\mathcal{M}(\mathcal{S}_{n}),\mathcal{Z}^{(t')}) - L_{emp}(\mathcal{M}(\mathcal{S}_{n}),\mathcal{S}_{n}^{(t')}) \Big| \leq \epsilon(\mathcal{S}^{(t')}_n) + 2B\sqrt{\frac{2K~\ln(2)~+~2~\ln(1/\mathcal{E})}{n^{(t')}}}~.$$
\end{lemma}

\begin{theorem}
\label{th: gen}
\textbf{(MALTS' distance metric is generalizable)}
The distance metric $\mathcal{M}(\cdot)$ learned using the data $\mathcal{S}_n$  and \textsc{MALTS} algorithm is generalizable and asymptotically generalizable, as follows:
\begin{enumerate}
    \item Generalizability: 
    
With probability at least 
$$(1-\mathcal{E})^{|\mathcal{T}|}\left(1-\exp\left(-\frac{\beta^2 \left(\rho^{(t')}_{\gamma}\right)^2}{ K^2 n^{(t')} B^2 }\right)\right)^{|\mathcal{T}|K^2}$$
with respect to the random draw of data,
\begin{eqnarray*}
\sum_{t'\in\mathcal{T}} \Big| L_{pop}(\mathcal{M}(\mathcal{S}_n),\mathcal{Z}^{(t')}) - L_{emp}(\mathcal{M}(\mathcal{S}_n),\mathcal{S}_n^{(t')}) \Big| 
     \leq
     2|\mathcal{T}|\beta+ \sum_{t'\in\mathcal{T}} 2B\sqrt{\frac{2K~\ln(2)~+~2~\ln(1/\mathcal{E})}{n^{(t')}}}
\end{eqnarray*}
for arbitrary chosen constants $\gamma>0$, $\mathcal{E}>0$, and $\beta\geq0$,
where $B$ is $\max_{z_1,z_2} loss(\mathcal{M}(\mathcal{S}_n),z_1,z_2)$, $\{C_i\}_{i=1}^{K}$ is the partition of $\mathcal{X}$ into non-empty sets $C_i$'s such that $K$ is the $\gamma$-covering number of $\mathcal{X}$, $C_i^{(t')} = \{z_j=(\x_j,y_j,t_j) : t_j=t',\x_j\in C_i \}$, and $\rho_{\gamma} = \min_{i,t'} |C^{(t')}_i|$.
\item Asymptotic Generalizability: $$ \hspace*{-25pt}\lim_{n\to\infty} \Bigg( \Big| L_{pop}(\mathcal{M}(\mathcal{S}_n),\mathcal{Z}^{(C)}) - L_{emp}(\mathcal{M}(\mathcal{S}_n),\mathcal{S}^{(C)}_n) \Big| + \Big| L_{pop}(\mathcal{M}(\mathcal{S}_n),\mathcal{Z}^{(T)}) - L_{emp}(\mathcal{M}(\mathcal{S}_n),\mathcal{S}^{(T)}_n) \Big| \Bigg) = 0$$
\end{enumerate}
\end{theorem}
Now that we have theoretically proven the functionality of MALTS, we will next discuss and compare MALTS performance with other methods on different datasets.
 	\section{Experiments}
\label{sec:Experiments}
In this section, we discuss and compare the performance of MALTS with other competing methods on a few different simulation setups with continuous covariates, discrete covariates and mixed (continuous and discrete) covariates. Lastly, we demonstrate MALTS performance for estimating ATE on LaLonde's NSW and PSID-2 data samples \citep{lalonde,dehejia_wahba_nonexp}. 

MALTS performs an $\eta$-fold honest causal inference procedure with the estimator $\phi$ inside each matched group being linear regression. We split the observed samples $\mathcal{S}_n$ into $\eta$ equal parts such that the ratio of treated to control units in each part is similar. For each fold, we use one of the $\eta$ partitions as the training set $\mathcal{S}_{tr}$ (not used for matching) and the rest of the $\eta-1$ partitions as the estimation set $\mathcal{S}_{est}$. 
Using the output from each of the $\eta$ folds, we calculate the estimated CATE for each unit (averaged across folds), estimated distance metric (averaged across folds) and a weighted unified matched group for each unit $s_i \in \mathcal{S}_n$. The weight of each matched unit $s_k$ corresponds to the number of times a particular unit $s_k$ was in the matched group of unit $s_i$ across the $\eta-1$ constructed matched groups. Here, $\eta$ was chosen to be 5 in our experiments. 

For interpretability, we let $\mathcal{M}_c$ be a diagonal matrix, which allows stretches of the continuous covariates. (Note that $\mathcal{M}_d$, which is the stretch matrix over discrete covariates, is always set to be diagonal.) This way, the magnitude of an entry in $\mathcal{M}_c$ or $\mathcal{M}_d$ provides the relative importance of the indicated covariate for the causal inference problem.

We further analyzed strategies for variance estimation for MALTS in Section~\ref{sec:coverage}, and performance under limited overlap between the covariates distribution of treated and control groups, and sensitivity to unobserved confounding. Detailed results are shown in \ref{sec:appendixB}.

The main results of these experiments are that \textbf{MALTS' performance is on par with existing state-of-the-art methods for causal inference}, including black box methods. \textbf{MALTS tends to have fairly consistent performance, even if the training set is fairly small or the number of irrelevant covariates is large}. Further, \textbf{MALTS provides interpretable distance metrics and matched groups} that black box machine learning methods do not provide.

\subsection{Data Generation Processes}
In this subsection we describe the data generation process (DGP) used in the simulation experiments. We use two main data-generation processes: The first DGP has a linear baseline with linear and quadratic treatment effects while the second DGP is the extension of Friedman's function introduced to test performance of prediction algorithms of \cite{friedman1991multivariate}. This second DGP, also termed as Friedman's DGP, has a scaled cosinusoidal treatment effect.
\subsubsection{Quadratic DGP} \label{dgp1}
This simulation includes both linear and quadratic terms.
Let $\mathbf{x}_{i,p} = \{\mathbf{x}_{i,p_c}, \mathbf{x}_{i,p_d}\}$ be a $p$-dimensional covariate vector composed of $|p_c|$ continuous covariates and $|p_d|$ discrete ones.
There are $k = k_{c} \cup k_{d}$  relevant covariates and the rest of the dimensions are irrelevant. Here, $p_c,k_c,p_d,$ and $k_d$ refer to the the subsets of indices of the covariates: all continuous, relevant continuous, all discrete, and relevant discrete, respectively.
$\mathbf{x}_{i,k_c}$ and $\mathbf{x}_{i,k_d}$ refer to the vectors of relevant continuous and discrete covariates respectively.  $\mathbf{x}_{i,k}$ refers to all $|k|$ relevant covariates. $\kappa_c \subseteq k_c$ is the set of continuous covariates and $\kappa_d \subseteq k_d$ is the set of discrete which are relevant in determining the treatment choice.
The potential outcomes and treatment assignment are determined as follows:
\begin{eqnarray*}
\lefteqn{
    \mathbf{x}_{i,p_c} \overset{iid}{\sim} \mathcal{N}(\mu,\Sigma), \
    \{x_{i,j}\}_{j\in p_d} \overset{iid}{\sim} \text{Bernoulli}(\psi), \
    \epsilon_{i,0},\epsilon_{i,1} \overset{iid}{\sim} \mathcal{N}(0,1), \ 
    \epsilon_{i,\textrm{treat}}\overset{iid}{\sim} \mathcal{N}(0,\sigma^2)}\\
    & \ s_1,\dots,s_{|k|} \overset{iid}{\sim} \text{Uniform}\{-1,1\}, \ \alpha_j|s_j \overset{iid}{\sim} \mathcal{N}(10s_j,9), \ \beta_1,\dots,\beta_{|k|} \overset{iid}{\sim} \mathcal{N}(1,0.25)
\end{eqnarray*}
\begin{eqnarray*}
y^{(0)}_i &=& \sum_{j\in k_c\cup k_d} \alpha_jx_{i,j} + \epsilon_{i,0}\\
y^{(1)}_i &=& \sum_{j\in k_c\cup k_d} \alpha_jx_{i,j} + \sum_{j\in k_c\cup k_d}\beta_jx_{i,j} + \sum_{j\in k_c \cup k_d} \sum_{j'\in k_c \cup k_d} x_{i,j}x_{i,j'} + \epsilon_{i,1}\\
t_i &=& \mathbbm{1}\left[\text{expit}\left(\sum_{j\in\kappa_c\subseteq k_c} x_{i,j} + \sum_{j\in\kappa_d\subseteq k_d} x_{i,j} - (|\kappa_c| \mu + |\kappa_d| \psi) +  \epsilon_{i,\textrm{treat}}\right)  > 0.5 \right] \\
y_i &=& t_i y^{(1)}_i + (1-t_i) y^{(0)}_i.
\end{eqnarray*}
Here expit$(z) = \exp(z)/(1+\exp(z))$. The variance of $\epsilon_{i,\textrm{treat}}$ determines how much confounding and overlap there is in the dataset: higher values of the variance make the dataset look like a randomized experiment with good overlap, while very small values of the variance lead to poor overlap and a very hard to analyze observational study. We explore these issues in detail in \ref{sec:appendixB}.

\subsubsection{Friedman's DGP} \label{dgp2}
The data generation process of \cite{friedman1991multivariate} was first proposed to assess the performance of prediction methods. We augmented Friedman's simulation setup to evaluate causal inference methods. The potential outcome under control is Friedman's function as provided by \citet{friedman1991multivariate} and \citet{Chipman10bart:bayesian}. The expected treatment effect we study is equal to the cosine of the product of the first two covariates scaled by the third covariate. 
\begin{eqnarray*}
\lefteqn{ x_{i,1} \dots x_{i,10} \overset{iid}{\sim} \mathcal{U}(0,1), \
    \epsilon_{i,0},\epsilon_{i,1} \sim \mathcal{N}(0,1), \ 
    \epsilon_{i,\textrm{treat}}\overset{iid}{\sim} \mathcal{N}(0,1)}\\
    y^{(0)}_i &=& 10~\sin(\pi x_{i,1} x_{i,2}) + 20~(x_{i,3} - 0.5)^2 + 10~x_{i,4} + 5~x_{i,5} + \epsilon_{i,0}\\
     y^{(1)}_i &=& 10~\sin(\pi x_{i,1} x_{i,2}) + 20~(x_{i,3} - 0.5)^2 + 10~x_{i,4} + 5~x_{i,5} + x_{i,3}~\cos(\pi x_{i,1} x_{i,2}) + \epsilon_{i,1}\\
     t_i &=& \mathbbm{1}\left[\text{expit}(x_{i,0} + x_{i,1} - 0.5 +  \epsilon_{i,\textrm{treat}})  > 0.5 \right] \\
    y_i &=& t_i y^{(1)}_i + (1-t_i) y^{(0)}_i.
\end{eqnarray*}
\subsection{Continuous Covariates}
\begin{figure}[ht]
     \centering
    \includegraphics[width=0.8\textwidth]{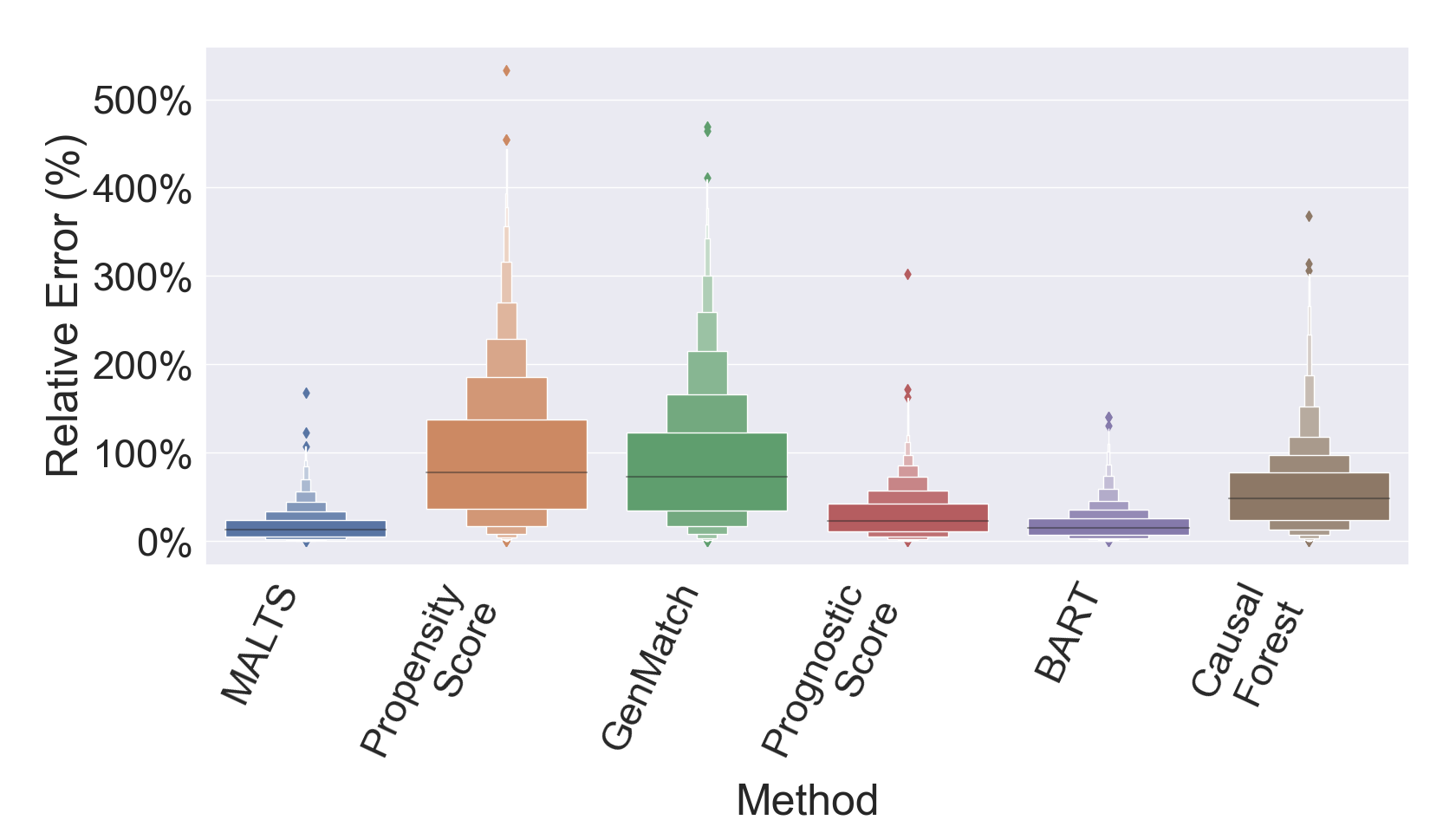}    \caption{{\textit{\it MALTS performs well with respect to other methods for continuous data.}} Letter-box plots of CATE Absolute Error relative to the true ATE on the test set for several methods.}
    \label{fig:continuous}
\end{figure}

We use the data-generation process described in Section~\ref{dgp1} to generate 2500 units with no discrete covariates, 15 important continuous covariates and 25 irrelevant continuous covariates. Further, we set the parameters for the DGP as follows:
$\mu = 1$, $\Sigma = 1.5 \textbf{I}$, $\psi=0.5$, $\sigma^2 = 1$ and $\kappa_c = \{0,1\}$.
We estimate CATE for each unit using matching methods like propensity score matching, prognostic score matching and genetic matching, and non-matching (uninterpretable) methods like causal forest and BART. Figure~\ref{fig:continuous} shows the performance of these methods. \textit{MALTS' performance is on par with existing state-of-the-art non-matching methods and outperforms all other matching methods for continuous covariates in the quadratic data generation process}. 

\subsection{Discrete Covariates}
We use the data-generation process described in Section~\ref{dgp1} to generate 2500 units with no continuous covariates, 15 important discrete covariates and 10 irrelevant discrete covariates. Further, we set the parameters of the DGP as follows: $\sigma^2 = 1$, $c = 2$ and $\kappa_d = \{0,1\}$. We used the weighted Hamming distance metric for this experiment.

\begin{figure}[ht]
     \centering
    \includegraphics[width=0.8\textwidth]{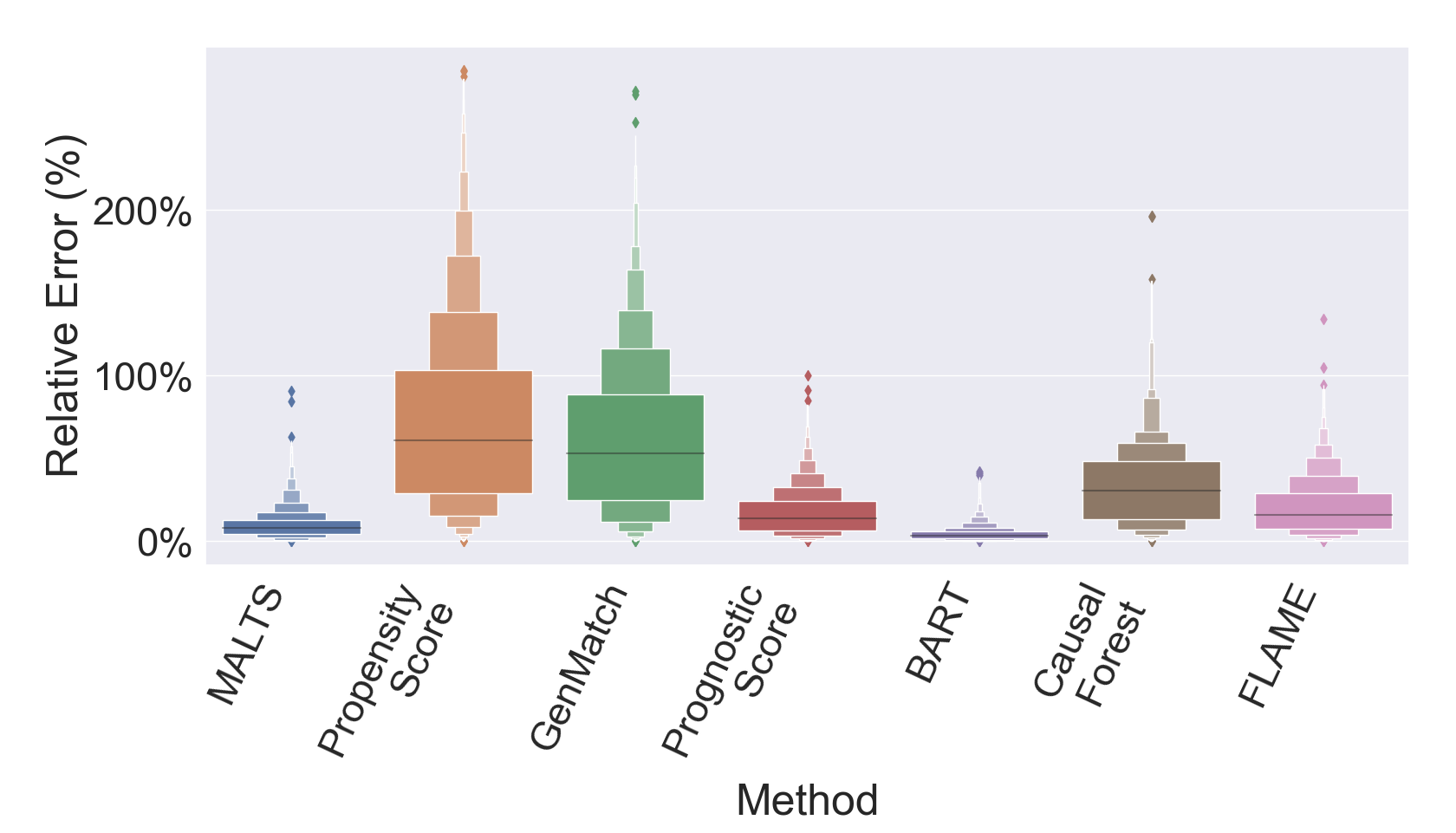}    \caption{{\textit{\it MALTS performs well with respect to other methods for discrete data.}} Letter-box plots of CATE Absolute Error relative to the true ATE on the test set for several methods.}
    \label{fig:discrete}
\end{figure}

Figure~\ref{fig:discrete} shows the performance comparison, again showing that 
MALTS' performance is on par with existing state-of-the-art non-matching methods; it also performs better than FLAME (a state-of-the-art matching method for discrete data) as it is able to provide additional smoothing in this relatively small-$n$ setting. \textit{Hence, MALTS performs well for discrete covariates in the quadratic data generation process.}

\subsection{Mixed Covariates}
\begin{figure}[ht]
    \centering
    \includegraphics[width=0.8\textwidth]{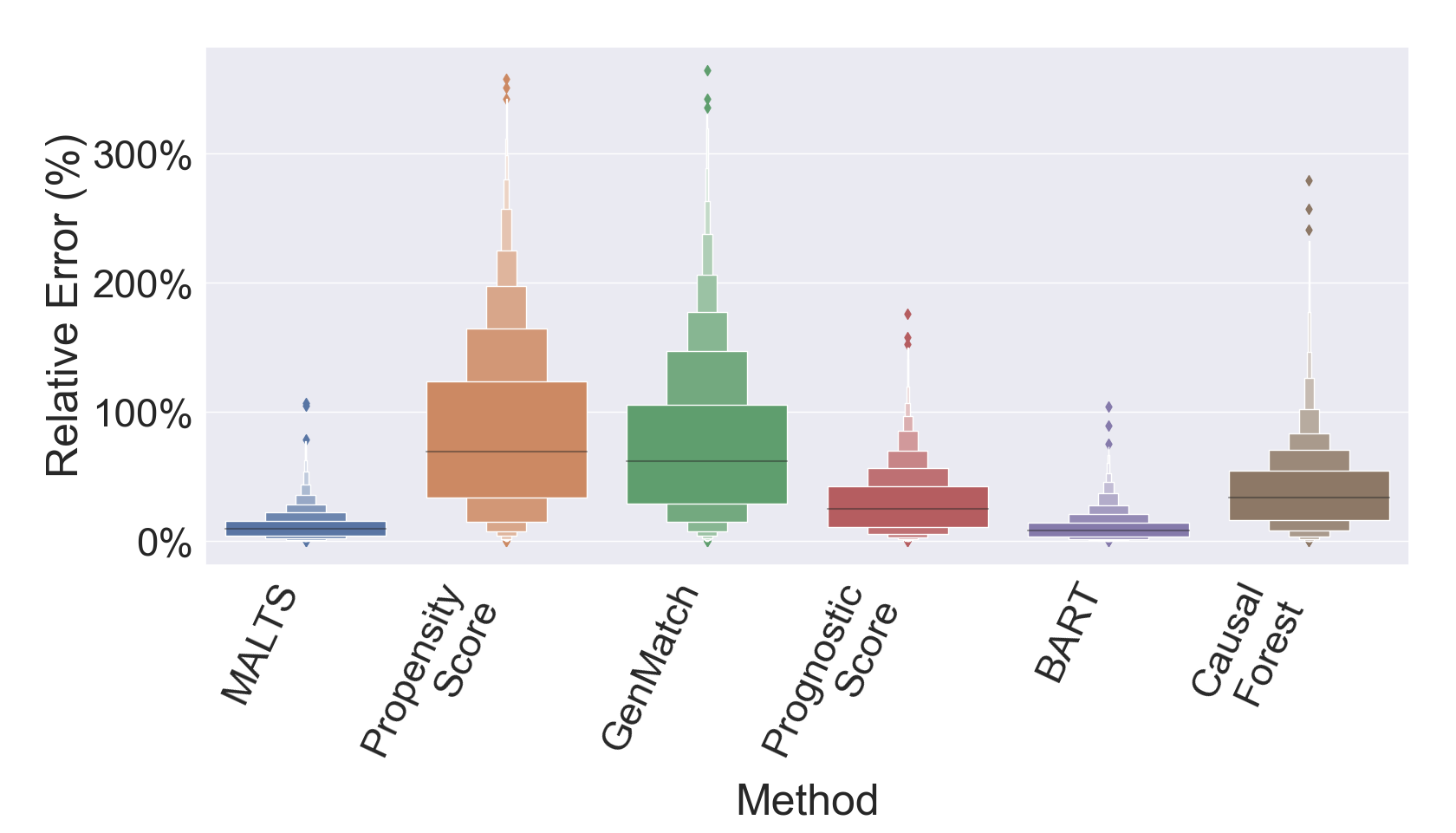}
    \caption{{\textit{\it MALTS performs well on data with mixed covariates.}} Letter-box plots of CATE Absolute Error relative to the true ATE on the test set for several methods. MALTS performs well on the setup with mixed (continuous+discrete) covariates.}
    \label{fig:mixed}
\end{figure}

We use the data-generation process used for experiments on continous and discrete covariates (described in Section~\ref{dgp1}) to generate 2500 units with 5 relevant continuous covariates, 15 relevant discrete covariates, 10 irrelevant continuous and 10 irrelevant discrete covariates. We used the same set of parameters for the DGP as the previous two experiments. Similar to the previous two experiments, Figure~\ref{fig:mixed} shows that \textit{MALTS performs on par with the state-of-the-art non-matching methods and outperforms all matching methods that can handle mixed covariates for the quadratic data generation process.}

\subsection{Number of Covariates}
\begin{figure}
    \centering
    \includegraphics[width =0.8\textwidth]{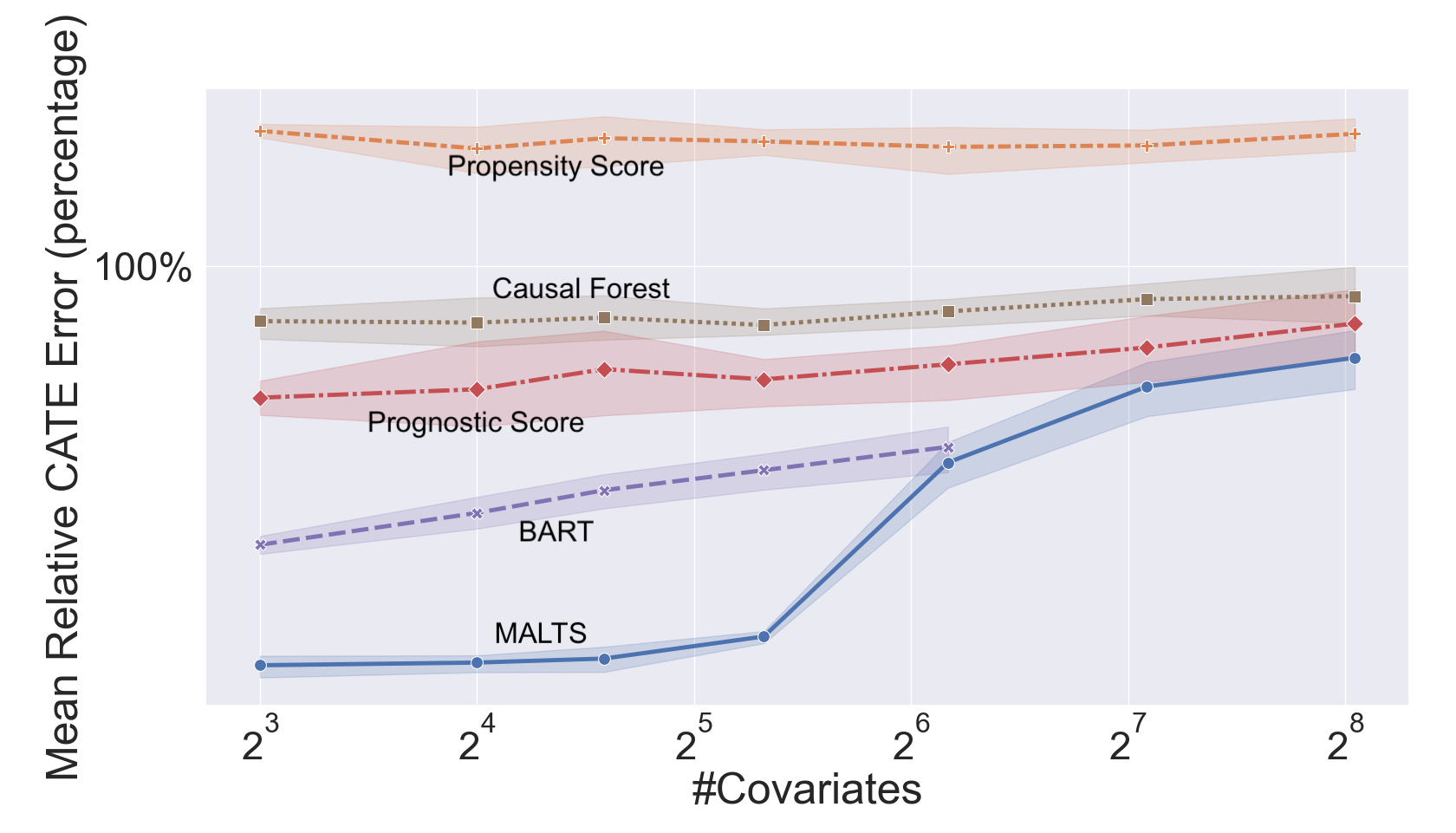}
    \caption{{\textit{\it MALTS performs on-par with other methods for a range of values of $p$.}} Comparative performance in estimating CATE using causal inference methods as the number of covariates increases, keeping the number of relevant covariates constant and equal to $8$. The number of units is fixed: $n = 2^{11}$. (For the given $n$, BART does not return CATE estimates for some units when $p>2^6$. Prognostic scores use BART for $p\leq 2^7$ and gradient boosted trees for $p>2^6$.)}
    \label{fig:kp}
\end{figure}

We studied the performance of various causal inference methods to estimate CATEs as the number of covariates ($p$) changes, keeping the number of relevant covariates ($|k|$) constant and equal to $8$. We simulated the data using the DGP described in Section~\ref{dgp1}. The number of units is constant ($n=2048$) while the number of covariates ($p$) changes from 8 to 256. The performance of MALTS is on-par with or better than other causal inference methods as the number of irrelevant covariates increases (see Figure~\ref{fig:kp}). This indicates that MALTS can be used to help reduce the effects of the curse of dimensionality.
\subsection{Number of Units}
\begin{figure}[ht]
    \centering
    \includegraphics[width=0.8\textwidth]{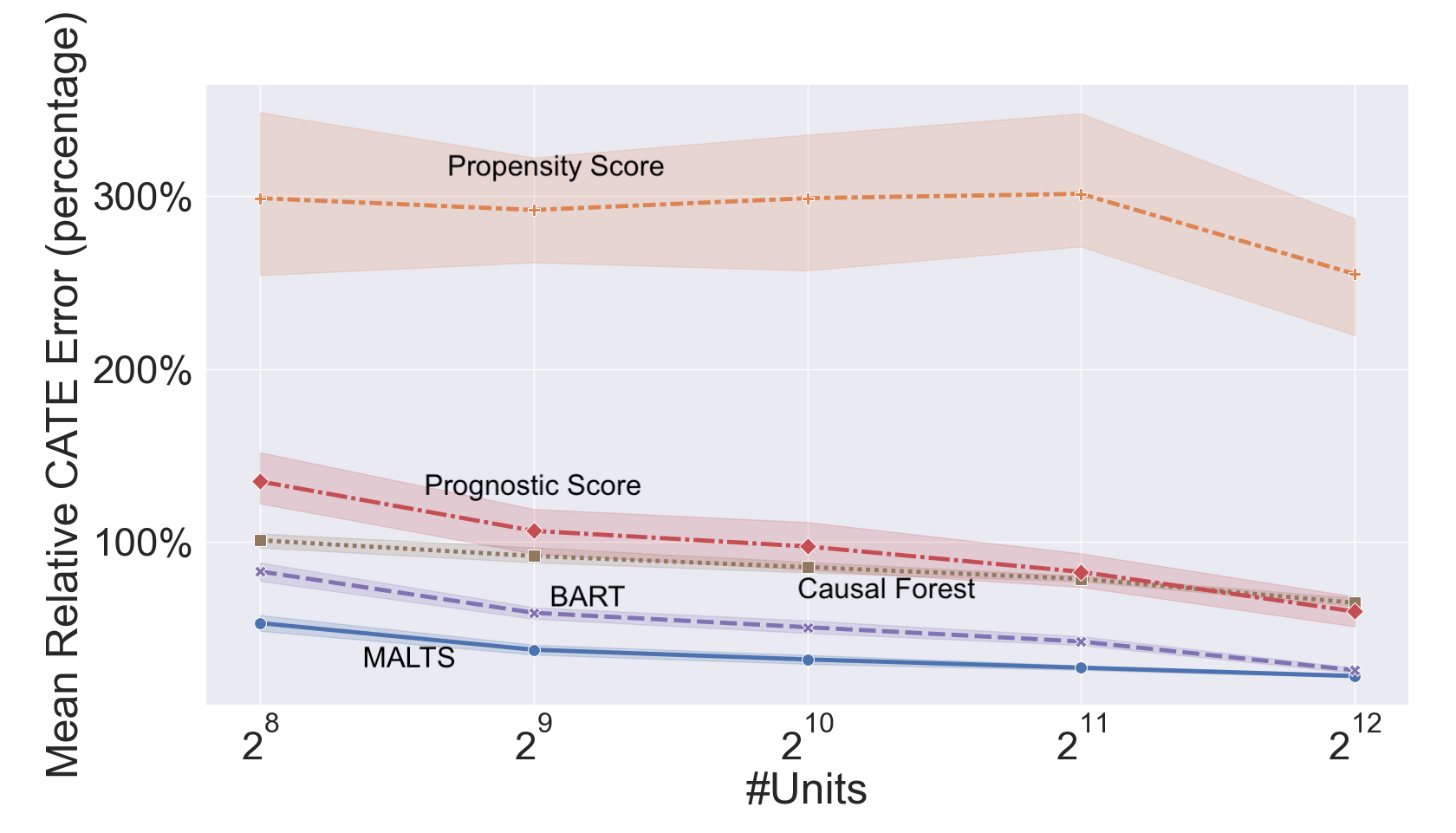}
    \caption{{\textit{\it MALTS consistently performs on par with or better than  non-interpretable approaches.}} Trend plots of average CATE Absolute Error for several methods, for different numbers of units in the datasets. }
    \label{fig:trend_n}
\end{figure}

We studied the change in CATE estimation error-rates as the number of units in a dataset increases. We simulated the data using the DGP described in Section~\ref{dgp1}, keeping the number of covariates constant and equal to $20$ (all of them are relevant in outcome determination). We changed the number of units from $2^8$ to $2^{12}$. \textit{MALTS' performance is on-par with or better than BART and the error-rate is significantly lower than that of other causal inference methods} (see Figure~\ref{fig:trend_n}).

\subsection{Friedman's Setup}
We further compare MALTS and other flexible methods' performance on data generated using the process described in Section~\ref{dgp2}. 
This DGP is particularly interesting because the potential outcomes are highly non-linear functions with trigonometric expressions.
\begin{figure}[ht]
    \centering
    \includegraphics[width=0.8\textwidth]{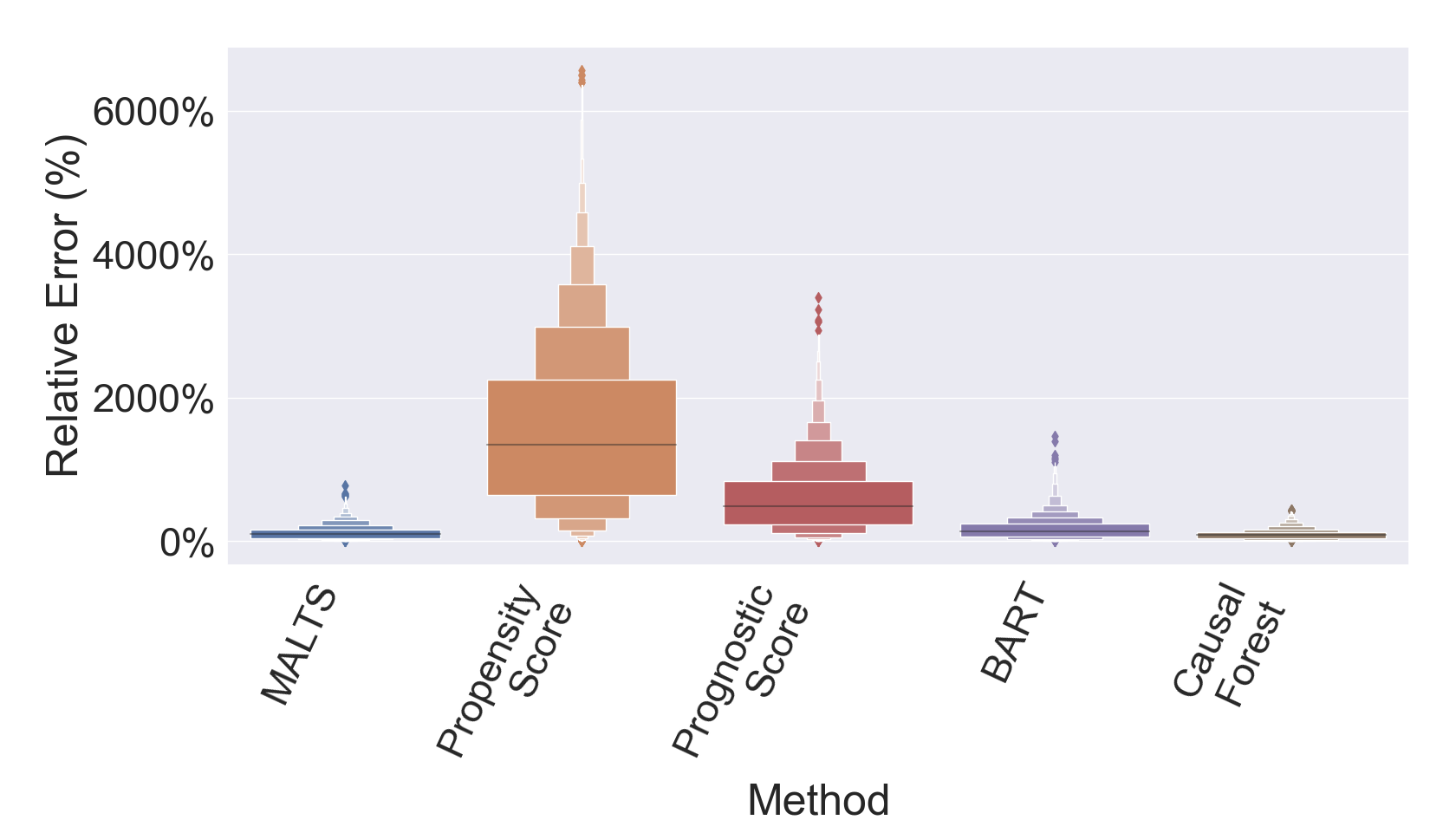}
    \caption{{\textit{\it MALTS performs well on Friedman's setup.}} Letter-box plots of CATE absolute error relative to true ATE for MALTS and other causal inference methods. 
}
    \label{fig:Friedman}
\end{figure}

As shown in Figure~\ref{fig:Friedman}, we observe that \textit{MALTS performs on par with Causal Forest while BART's error-rate is significantly higher (worse) than MALTS, for the Friedman's data generation process}.

\subsection{Coverage Study}\label{sec:coverage}
\begin{figure}
    \centering
    \includegraphics[width=\textwidth]{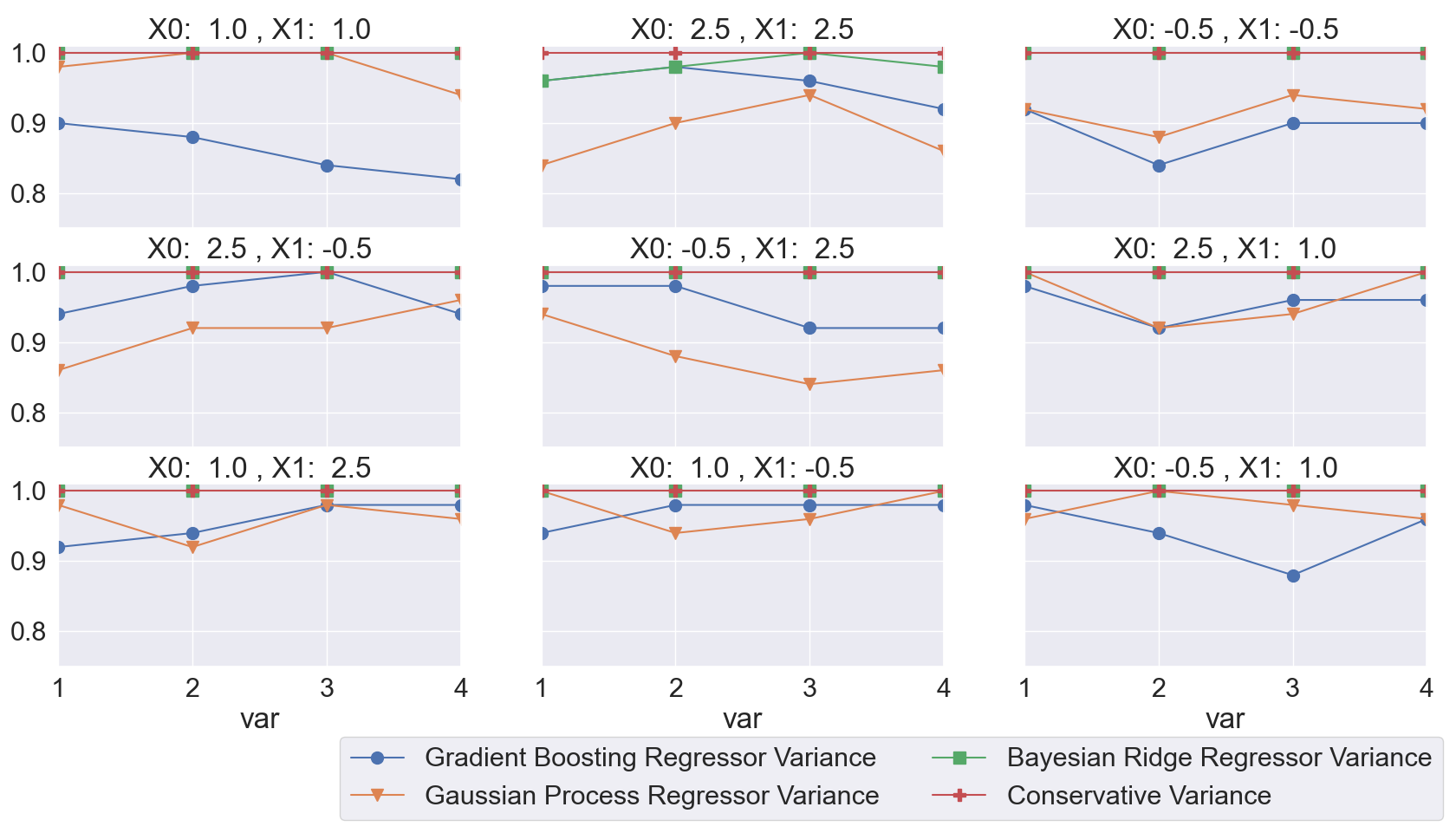}\\(a)\\
    \includegraphics[width=0.7\textwidth]{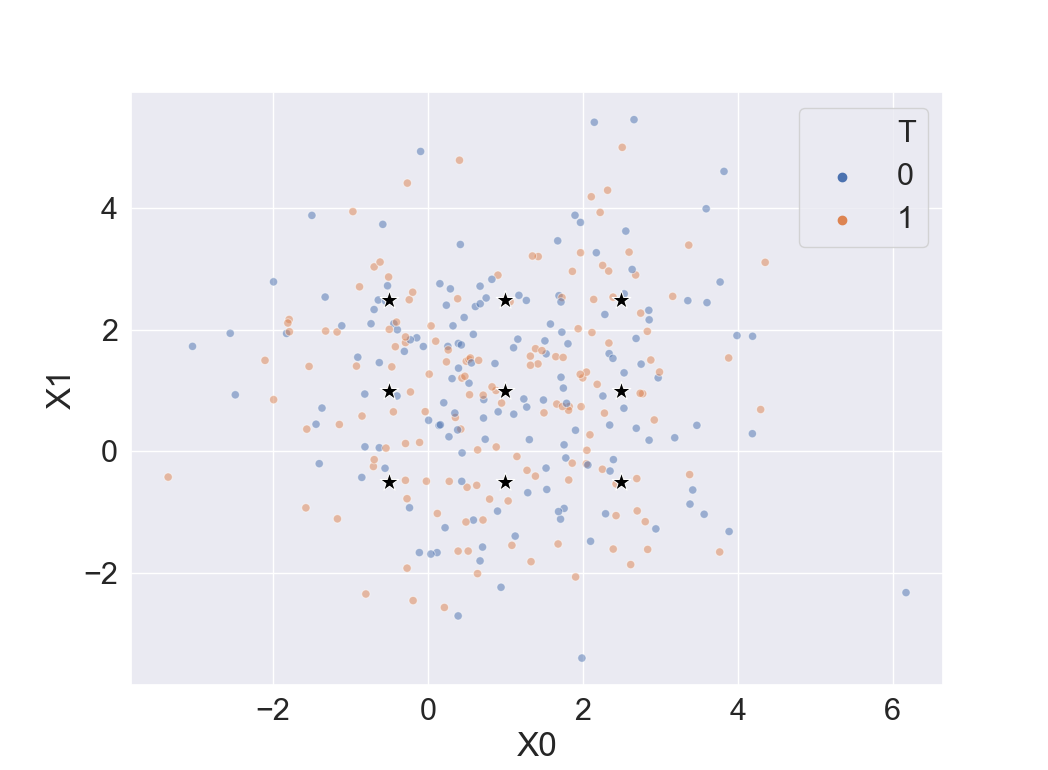}\\(b)
    \caption{(a) Coverage of 95 percent confidence interval for 9 points: (1.0,1.0), (2.5,2.5), (-0.5,-0.5), (2.5,-0.5), (-0.5,2.5), (4.0,4.0), (-3.0,-3.0), (4.0,-3.0) and (-3.0,4.0). (b) Covariate space showing positions of 9 points-of-interest as black-stars, with other points color-coded according to their treatment assignments.}
    \label{fig:Coverage}
\end{figure}

We use the DGP described in Section~\ref{dgp1} with 2 relevant continuous covariates and no irrelevant covariates for the coverage study. Further, we set the parameters to the DGP as follows:
$\mu = 1$, $\Sigma = 1.5 \textbf{I}$, $\psi=0.5$, and $c = 2$.
We selected 9 reference points in a grid from the covariate space as shown in Figure~\ref{fig:Coverage}(b) and conducted an experiment that considered these reference points, over 100 repetitions. We compared coverage for CATEs estimated using MALTS for different values of the variance, ranging from 1.0 to 4.0, for noise term $\epsilon_0$ and $\epsilon_1$ in the potential outcomes function. 

Variance estimation is notoriously hard in matching problems, even for overall quantities such as the average treatment effect \citep{abadie2006}. We consider both a conservative variance estimator \citep{wang2017flame} and estimators that sacrifice some interpretability for better coverage. Specifically, we consider the CATEs estimated using MALTS and study how well an uninterpretable method can predict those estimates to obtain a variance estimate. We use the predictive variance from gradient boosting regression, from gaussian process regression and from Bayesian ridge regression on the covariates, where we estimated CATEs and quantify variance of each CATE estimate.

Based on Figure~\ref{fig:Coverage}(a), the coverage for each the nine points of interest is between 0.85 and 1 for most values of the variance using any of the three variance estimation approaches.

\subsection{LaLonde Data}
The LaLonde data pertain to the National Support Work Demonstration (NSW) temporary employment program and its effect on income level of the participants \citep{lalonde}. This dataset is frequently used as a benchmark for the performance of methods for observational causal inference. We employ the male sub-sample from the NSW in our analysis as well as the PSID-2 control sample of male household-heads under age 55 who did not classify themselves as retired in 1975 and who were not working when surveyed in the spring of 1976 \citep{dehejia_wahba_nonexp}. The outcome variable for both experimental and observational analyses is earnings in 1978 and the considered variables are age, education, whether a respondent is Black, is Hispanic, is married, has a degree, and their earnings in 1975. Previously, it has been demonstrated that almost any adjustment during the analysis of the experimental and observational variants of these data (both by modeling the outcome and by modeling the treatment variable) can lead to extreme bias in the estimate of average treatment effects \citep{lalonde}.

\begin{table}[ht]
 \caption{{\it Estimated ATE for different methods on Lalonde's NSW experimental dataset. The MALTS estimate of ATE is closer to the true ATE than other methods. We provide estimates for MALTS before and after pruning the matched groups with large diameters. The threshold to prune was chosen by rule of thumb on diameters of matched groups as shown in Figure~\ref{fig:lalonde_prune}(b).}}
 \label{tab:lalonde}
 \centering
\begin{tabular}{lrr}
\hline
{} &  \textbf{ATE Estimate} &  \textbf{Estimation Bias (\%)} \\
\textbf{Method}           &               &                      \\
\hline
Truth         &    886 &            - \\
\textit{MALTS}            &    \textit{881.67} &            \textit{-0.49} \\
\textit{MALTS  (pruned)}            &    \textit{888.53} &            \textit{0.29} \\
GenMatch         &    859.72 &            -2.97 \\
Propensity Score &    513.30 &           -42.06 \\
Prognostic Score &    943.81 &             6.52 \\
BART-CV          &   1164.72 &            31.46 \\
Causal Forest-CV &    509.32 &           -42.51 \\
\hline
\end{tabular}
\end{table}

\begin{table}[ht]
 \caption{{\it Estimated ATE for different methods on Lalonde's NSW experimental data and PSID-2 observational dataset.} We provide estimates for MALTS before and after pruning the matched groups with large diameters. The threshold to prune was chosen by rule of thumb on diameters of matched groups as shown in Figure~\ref{fig:lalonde_prune}(b).}
 \label{tab:lalonde_psid2}
 \centering
\begin{tabular}{lrr}
\hline
{} &  \textbf{ATE Estimate} &  \textbf{Estimation Bias (\%)} \\
\textbf{Method}          &               &                      \\
\hline
Truth         &    886 &            - \\
\textit{MALTS}            &    \textit{608.37} &             \textit{-31.34} \\
\textit{MALTS (pruned)}            &    \textit{891.75} &             \textit{0.65} \\
GenMatch         &    549.53 &           -37.98 \\
Propensity Score &    513.79 &           -42.01 \\
Prognostic Score &   -897.76 &          -201.33 \\
BART-CV          &    713.20 &           -19.50 \\
Causal Forest-CV &   -179.98 &          -120.31 \\
\hline
\end{tabular}
\end{table}

\textbf{Performance results:} Tables~\ref{tab:lalonde} and \ref{tab:lalonde_psid2} present the average treatment effect estimates based on MALTS, state-of-the-art modeling methods, and matching methods. \textit{MALTS (after appropriately pruning low-quality matched groups) is able to achieve accurate ATE estimation on both experimental and observational datasets.}

\begin{figure}[ht]
    \centering
    \includegraphics[width=0.75\textwidth]{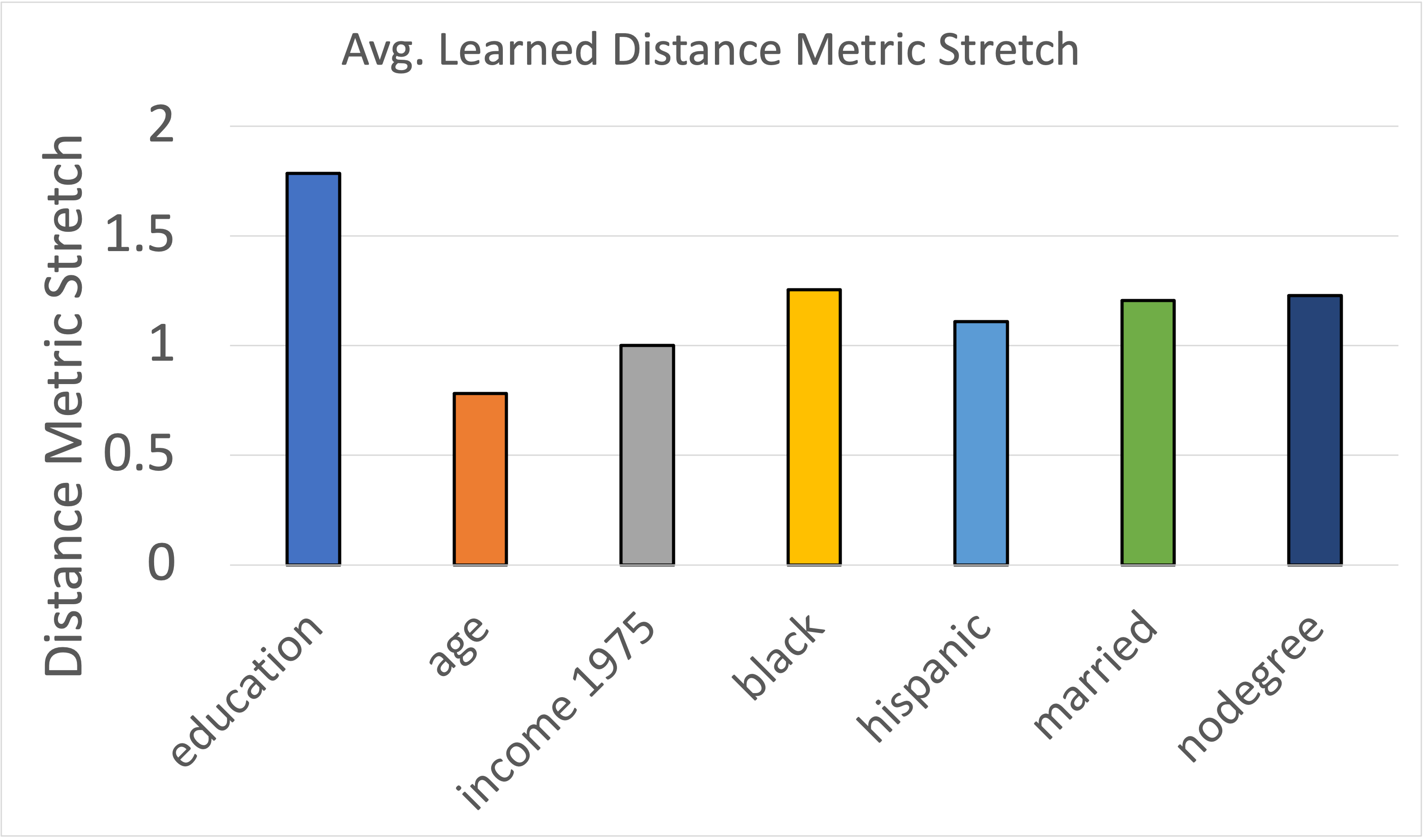}\\
    (a) Distance metric learned on Lalonde data across 250 folds -- 50 repeats and 5 splits within each repeat. Here, on an \textit{average}, education is stretched more than other variables, which means it is more important to match closely on education.\\
    \includegraphics[width=0.8\textwidth]{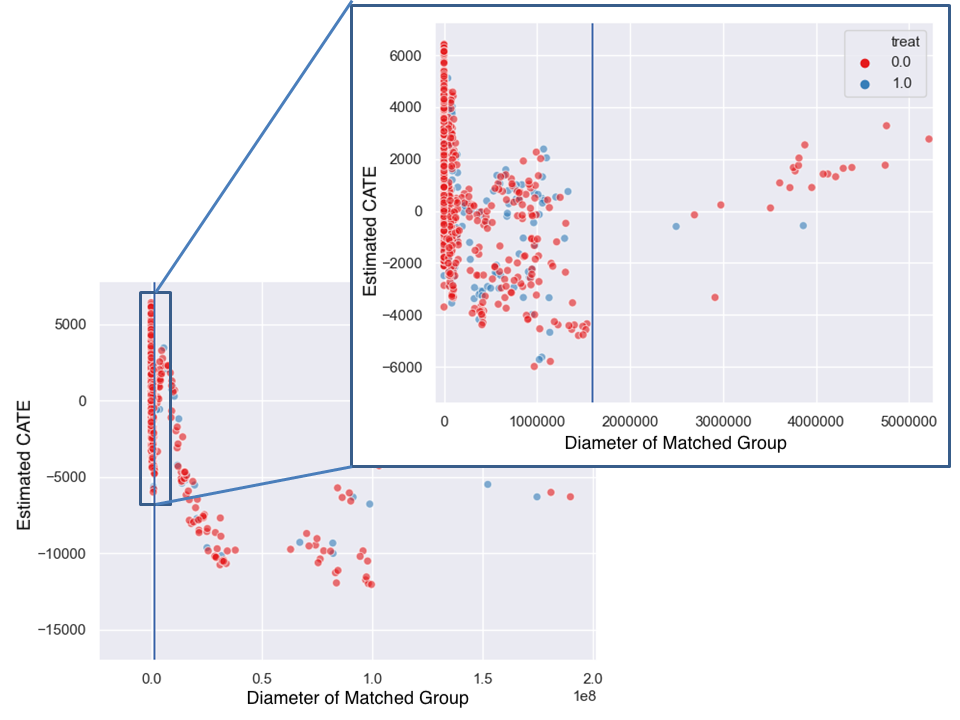}\\
    (b) Criteria for pruning low-quality matched groups with large diameter.\\
    \caption{(a) Box-plot of distance metric stretch values corresponding to each covariate in Lalonde data learned over 5 folds and 20 repeats.  (b) Criteria to prune low-quality matched groups with large diameter from Lalonde data.}
    \label{fig:lalonde_prune}
\end{figure}

Figure~\ref{fig:lalonde_prune} illustrates how the matched groups were pruned. There was a clear visual separation between high-quality matched groups, which had low diameters, and low-quality matched groups, with larger diameters.

\textbf{Model Interpretability:}
One difference between MALTS and the other methods is that its solution can be described concisely: MALTS produces a total of seven numbers that define the distance metric on the LaLonde data. The distribution of the learned distance metric values across folds is shown in Figure~\ref{fig:lalonde_prune}(a). Once the researcher has these seven numbers, along with the value of $k$ in $k$-nearest neighbors used to train MALTS, they know precisely which units should be matched. In contrast, causal forest and BART require a model whose size  depends on the number of trees, where each tree is several levels deep--in this case, 2000 trees  and 150 trees, respectively.

\textbf{Interpretability of Matched Groups:}
To examine the interpretability of MALTS' matched groups, we present two of the matched groups from MALTS for the observational Lalonde dataset in Table~\ref{tab:match-group-example}, corresponding to two ``query'' individuals in the dataset.
Query 1 is a 22 year old with no income in 1975. MALTS was able to construct a tight matched group for this individual (both in control and in treatment). In contrast, Query 2 is a 42-year-old high-income individual without a degree, which is an extremely unlikely scenario, leading to a matched group with a very large diameter, which should probably not be used during analysis. Such granular analysis is not possible for regression methods like BART and matching methods like prognostic score or propensity score matching.

 This further highlights the troubleshooting capabilities of interpretable matching methods: by identifying units that are poorly matched, we know exactly which units to study in more detail. In this case, it is possible that the ``degree'' field might have a data error, which means it would be better not to match this unit and to potentially follow up on the veracity of responses to the survey.

\begin{table}[ht]
\caption{{\it Learned distance metric and examples of matched-groups on Lalonde Experimental treatment and Observational control datasets for two example query points drawn from the same datasets.} Query 1 represents a high quality (low diameter) matched group while Query 2 represents a poor quality (high diameter) matched group that could be discarded during analysis.}
\label{tab:match-group-example}
\center{\textbf{Stretch Matrix}}\\
\resizebox{\textwidth}{!}{\begin{tabular}{rrrrrrrrrr}
\hline
\textbf{} &
  \textbf{} &
  \textbf{Age} &
  \textbf{Education} &
  \textbf{Black} &
  \textbf{Hispanic} &
  \textbf{Married} &
  \textbf{No-Degree} &
  \textbf{Income-1975} &
  \textbf{} \\ \hline
mean($Diag(\mathcal{M})$) & & 0.780 & 1.786 & 1.254 & 1.110 & 1.205 & 1.229 & 1.001 &  \\
std($Diag(\mathcal{M})$) & & 0.361 & 0.778 & 0.641 & 0.577 & 0.614 & 0.618 & 0.512 &  \\ \hline
\end{tabular}}
\vskip 5mm
\center{\textbf{Two Matched Groups}}
\resizebox{\textwidth}{!}{\begin{tabular}{lrrrrrrrrr}
\hline
\textbf{Unit-ID} &
  \textbf{Treated} &
  \textbf{Age} &
  \textbf{Education} &
  \textbf{Black} &
  \textbf{Hispanic} &
  \textbf{Married} &
  \textbf{No-Degree} &
  \textbf{Income-1975} &
  \textbf{Income-1978} \\ \hline
Query-1: \textit{1}   & Yes & 22 & 9  & No & Yes & No & Yes & \$0 & \$3595  \\  \hline
        \textit{94}  & Yes & 23 & 8  & No & Yes & No & Yes & \$0 & \$3881  \\
        \textit{330} & No & 22 & 8  & No & Yes & No & Yes & \$0 & \$9920  \\
        \textit{299} & No & 22 & 9  & Yes & No & No & Yes & \$0 & \$0     \\
        \textit{5}   & Yes & 22 & 9  & Yes & No & No & Yes & \$0 & \$4056  \\
        \textit{82}  & Yes & 21 & 9  & Yes & No & No & Yes & \$0 & \$0    \\
        \textit{416} & No & 22 & 9  & Yes & No & No & Yes & \$0 & \$12898 \\
        \textit{333} & No & 21 & 9  & Yes & No & No & Yes & \$0 & \$3343  \\
        \textit{292} & Yes & 20 & 9  & Yes & No & No & Yes & \$0 & \$8881  \\
        \textit{17}  & Yes & 23 & 10 & Yes & No & No & Yes & \$0 & \$7693  \\
        \textit{116} & Yes & 24 & 10 & Yes & No & No & Yes & \$0 & \$0    \\ \hline
\end{tabular}}
\vskip 5mm
\resizebox{\textwidth}{!}{\begin{tabular}{lrrrrrrrrr}
\hline
\textbf{Unit-ID} &
  \textbf{Treated} &
  \textbf{Age} &
  \textbf{Education} &
  \textbf{Black} &
  \textbf{Hispanic} &
  \textbf{Married} &
  \textbf{No-Degree} &
  \textbf{Income-1975} &
  \textbf{Income-1978} \\
\hline
Query-2: \textit{968} &    No &  42 &       11 &    No &       No &      Yes &       Yes &  \$44758 &  \$54675 \\
\hline
\textit{274} &    Yes &  35 &        9 &    Yes &       No &      Yes &       Yes &  \$13830 &  \$12803 \\
\textit{141} &    Yes &  25 &        8 &    Yes &       No &      No &       Yes &  \$37431 &   \$2346 \\
\textit{967} &    No &  50 &       17 &    No &       No &      Yes &       No &  \$30435 &  \$25860 \\
\textit{948} &    No &  35 &       12 &    No &       No &      Yes &       No &  \$26854 &  \$29554 \\
\textit{210} &    Yes &  25 &        8 &    No &       No &      No &       Yes &  \$23096 &   \$6421 \\
\textit{241} &    Yes &  24 &       15 &    Yes &       No &      No &       No &  \$13008 &  \$14683 \\
\textit{311} &    No &  28 &       12 &    Yes &       No &      Yes &       No &  \$29009 &  \$10067 \\
\textit{183} &    Yes &  23 &       10 &    Yes &       No &      No &       Yes &  \$15709 &   \$5665 \\
\textit{182} &    Yes &  23 &       12 &    Yes &       No &      Yes &       No &  \$15079 &  \$10283 \\
\hline
\end{tabular}
}
\end{table} 	\section{Conclusion and Discussion}\label{sec:Conclusion}
This paper introduces the MALTS algorithm, which learns a distance metric on the covariate space for use with matching. The learned metric stretches important covariates and compresses irrelevant covariates for outcome prediction in order to produce high-quality matches. Unlike other methods, MALTS can handle a large number of irrelevant covariates by compressing them to the point where they are effectively eliminated, which helps handle the curse of dimensionality.
Unlike black-box machine learning methods, MALTS produces interpretable matched groups and returns the stretch matrix on covariates for counterfactual prediction. The stretch matrix is chosen here to be diagonal, so that it can be represented using only a few ``stretch'' numbers that determine the importance of each covariate in determining the matched groups. 

Whereas deep neural networks mainly show improvements over other methods for problems that do not have natural data representations (computer vision, speech, etc.), we conjecture that the stretch/almost-exact match combination should suffice for most datasets. A natural extension, however, is to use neural networks to learn a flexible distance metric in a latent space, thus allowing us to match on medical records, images, and text documents.  This will allow us to incorporate complex data structures by introducing a flexible learning framework (e.g., interpretable neural networks) for coding the data. That is, we can redefine the distance metric via
\begin{eqnarray*}\nonumber
\dis_{\mathcal{M}} (\x_i,\x_j) = \langle\omega_{\mathcal{M}}(\x_i),\omega_{\mathcal{M}}(\x_j) \rangle \;\;\;\textrm{or} \;\;\;\nonumber
\dis_{\mathcal{M}} (\x_i,\x_j) = \left(\omega_{\mathcal{M}}(\x_i)-\omega_{\mathcal{M}}(\x_j)\right)^2,
\end{eqnarray*}
where $\omega_\mathcal{M}$ is a summary of relevant data features learned using a complex modeling framework. 

In the future, the MALTS framework could be extended to deal with missing covariates, and can be adapted to instrumental variables. 
 	
\acks{We gratefully acknowledge funding from the National Science Foundation under grants III 1703431, CCF 1934964, IIS 2130250, IIS 2147061 (with Amazon), and CAREER DMS 2046880, and the National Institute of Health under grants NIDA DA054994 and R01EB025021. We also acknowledge funding from an Amazon Graduate fellowship.}

\bibliography{biblio}{}
\appendix
\renewcommand{\thesection}{Appendix \Alph{section}}
    \section{}\label{sec:appendixA}
In this section we provide proofs for theorems and lemmas discussed in Section~\ref{sec:theory}.\\
\allowdisplaybreaks

\noindent \textbf{Proof (Lemma~\ref{lm: whpavgloss})}. If $(D_1,\dots,D_K)$ is the multinomially distributed random vector with parameters $d$ and $p_1, \dots, p_K$ then, by the Bretagnolle-Huber-Carol inequality, $$ P\left(\sum_{i=1}^{K} \Big| \frac{D_i}{d} - p_i \Big|\geq\lambda\right)\leq 2^K e^{-\frac{d\lambda^2}{2}}.$$ Thus, for our case, we can consider $N_i$ corresponding to the set of indices of units in sample $\mathcal{S}^{(t')}_n$ such that their $x$'s are contained in the partition $\mathbf{C}_{i}$ as in Theorem~\ref{th: robust}. Hence, by the Bretagnolle-Huber-Carol inequality, we know that 
$$
    P\Bigg(~ \sum_{i=1}^{K} \Big| \frac{|N_i|}{n^{(t')}} - \mu(\mathbf{C}_{i}) \Big| \geq \sqrt{\frac{2K~\ln(2)~+~2~\ln(1/\mathcal{E})}{n^{(t')}}} ~\Bigg) \leq \mathcal{E}~.
$$
Now, for some arbitrary $t'\in\mathcal{T}$ let us consider $\Big| L_{pop}(\mathcal{M}(\mathcal{S}_n),\mathcal{Z}^{(t')}) - L_{emp}(\mathcal{M}(\mathcal{S}_n),\mathcal{S}_n^{(t')}) \Big|$. 
We know that
{\allowdisplaybreaks
\begin{eqnarray*}
\lefteqn{
        \Big| L_{pop}(\mathcal{M}(\mathcal{S}_n),\mathcal{Z}^{(t')}) - L_{emp}(\mathcal{M}(\mathcal{S}_n),\mathcal{S}_n^{(t')}) \Big| }\\
       & =& 
        \Bigg| \sum_{i,j=1}^{K} \Big( \mathbb{E}_{z_1,z_2}[loss(\mathcal{M}(\mathcal{S}_n),z_1=(\x'_1,y'_1,t'_1),z_2=(\x'_2,y'_2,t'_2))~|~\x'_1\in\mathbf{C}_i,\x'_2\in\mathbf{C}_j]~\mu(\mathbf{C}_i)\mu(\mathbf{C}_j) \Big)
        \\&&- 
        \frac{1}{(n^{(t')})^2}\sum_{s_1,s_2\in\mathcal{S}^{(t')}_n} loss(\mathcal{M}(\mathcal{S}_n),s_1,s_2)
        \Bigg|\\
        & =& 
        \Bigg| \sum_{i,j=1}^{K} \left( \mathbb{E}_{z_1,z_2}[loss(\mathcal{M}(\mathcal{S}_n),z_1,z_2)~|~\x'_1\in\mathbf{C}_i,\x'_2\in\mathbf{C}_j]~\mu(\mathbf{C}_i)\mu(\mathbf{C}_j) \right) \\&& - \sum_{i,j=1}^{K} \left(\mathbb{E}_{z_1,z_2}[loss(\mathcal{M}(\mathcal{S}_n),z_1,z_2)~|~\x'_1\in\mathbf{C}_i,\x'_2\in\mathbf{C}_j]~\mu(\mathbf{C}_i)~\frac{|N_j|}{n^{(t')}}\right)
        \\&&
        + \sum_{i,j=1}^{K} \left(\mathbb{E}_{z_1,z_2}[loss(\mathcal{M}(\mathcal{S}_n),z_1,z_2)~|~\x'_1\in\mathbf{C}_i,\x'_2\in\mathbf{C}_j]~\mu(\mathbf{C}_i)~\frac{|N_j|}{n^{(t')}}\right) \\&&+ \sum_{i,j=1}^{K} \mathbb{E}_{z_1,z_2}[loss(\mathcal{M}(\mathcal{S}_n),z_1,z_2)~|~\x'_1\in\mathbf{C}_i,\x'_2\in\mathbf{C}_j]~
        \frac{|N_i|}{n^{(t')}} \frac{|N_j|}{n^{(t')}} 
        \\&&- \sum_{i,j=1}^{K} \mathbb{E}_{z_1,z_2}[loss(\mathcal{M}(\mathcal{S}_n),z_1,z_2)~|~\x'_1\in\mathbf{C}_i,\x'_2\in\mathbf{C}_j]~
        \frac{|N_i|}{n^{(t')}} \frac{|N_j|}{n^{(t')}}
        \\&&- 
        \frac{1}{(n^{(t')})^2}\sum_{s_1,s_2\in\mathcal{S}^{(t')}_n} loss(\mathcal{M}(\mathcal{S}_n),s_1,s_2)
        \Bigg|\\
        &\leq& \Bigg| \sum_{i,j=1}^{K} \mathbb{E}_{z_1,z_2}[loss(\mathcal{M}(\mathcal{S}_n),z_1,z_2)~|~\x'_1\in\mathbf{C}_i,\x'_2\in\mathbf{C}_j]~\mu(\mathbf{C}_i)\Big(\mu(\mathbf{C}_j) - \frac{|N_j|}{n^{(t')}}\Big) \Bigg|\\
        &&+\Bigg| \sum_{i,j=1}^{K} \mathbb{E}_{z_1,z_2}[loss(\mathcal{M}(\mathcal{S}_n),z_1,z_2)~|~\x'_1\in\mathbf{C}_i,\x'_2\in\mathbf{C}_j]~
        \frac{|N_j|}{n^{(t')}}
        \Big(\mu(\mathbf{C}_i) - \frac{|N_i|}{n^{(t')}}\Big) \Bigg|\\
       && + \Bigg| \sum_{i,j=1}^{K} \mathbb{E}_{z_1,z_2}[loss(\mathcal{M}(\mathcal{S}_n),z_1,z_2)~|~\x'_1\in\mathbf{C}_i,\x'_2\in\mathbf{C}_j]~
        \frac{|N_i|}{n^{(t')}} \frac{|N_j|}{n^{(t')}} \\&& - \frac{1}{(n^{(t')})^2}\sum_{s_1,s_2\in\mathcal{S}^{(t')}_n} loss(\mathcal{M}(\mathcal{S}_n),s_1,s_2) \Bigg|\\
    &\leq& 2B\sum_{i=1}^{K}\Big| \frac{|N_i|}{n^{(t')}} - \mu(\mathbf{C}_i) \Big| + \Bigg| \sum_{i,j=1}^{K} \mathbb{E}_{z_1,z_2}[loss(\mathcal{M}(\mathcal{S}_n),z_1,z_2)~|~\x'_1\in\mathbf{C}_i,\x'_2\in\mathbf{C}_j]~
        \frac{|N_i|}{n^{(t')}} \frac{|N_j|}{n^{(t')}} \\&& - \frac{1}{(n^{(t')})^2}\sum_{s_1,s_2\in\mathcal{S}^{(t')}_n} loss(\mathcal{M}(\mathcal{S}_n),s_1,s_2) \Bigg| \text{ where $B$ is }\max_{z_1,z_2} loss(\mathcal{M}(\mathcal{S}_n),z_1,z_2).
\end{eqnarray*}
}
Recall, $\mathcal{M}(\mathcal{S}_n)$ is $(K,\epsilon(\cdot))$-multi-robust with probability $p_{mr}(\epsilon)$. Thus, 
\begin{eqnarray*}
&&P\Bigg( \begin{vmatrix} \sum_{i,j=1}^{K} \mathbb{E}_{z_1,z_2}[loss(\mathcal{M}(\mathcal{S}_n),z_1,z_2)~|~\x'_1\in\mathbf{C}_i,\x'_2\in\mathbf{C}_j]~
        \frac{|N_i|}{n^{(t')}} \frac{|N_j|}{n^{(t')}} \\ - \frac{1}{(n^{(t')})^2}\sum_{s_1,s_2\in\mathcal{S}^{(t')}_n} loss(\mathcal{M}(\mathcal{S}_n),s_1,s_2) \end{vmatrix} \leq \epsilon(\mathcal{S}_n^{(t')}) \Bigg) \\ 
        &&\geq \prod_{i,j} P\Bigg( \frac{|N_i||N_j|}{(n^{(t')})^2} \begin{vmatrix} \mathbb{E}_{z_1,z_2}[loss(\mathcal{M}(\mathcal{S}_n),z_1,z_2)~|~\x'_1\in\mathbf{C}_i,\x'_2\in\mathbf{C}_j] \\ - \frac{1}{|N_i||N_j|}\sum_{s_1,s_2\in\mathcal{S}^{(t')}_n} loss(\mathcal{M}(\mathcal{S}_n),s_1,s_2) \end{vmatrix} \leq \epsilon(\mathcal{S}_n^{(t')})/K^2 \Bigg)
        \\ 
        &&\geq \prod_{i,j} P\Bigg( \begin{vmatrix} \mathbb{E}_{z_1,z_2}[loss(\mathcal{M}(\mathcal{S}_n),z_1,z_2)~|~\x'_1\in\mathbf{C}_i,\x'_2\in\mathbf{C}_j] \\ - \frac{1}{|N_i||N_j|}\sum_{(s_1,s_2)\in(\mathbf{C}_i\times \mathbf{C}_j)} loss(\mathcal{M}(\mathcal{S}_n),s_1,s_2) \end{vmatrix} \leq \epsilon(\mathcal{S}_n^{(t')})/K^2 \Bigg)
        \\&&  \geq  (p_{mr}(\epsilon/K^2))^{K^2}
\end{eqnarray*}
Hence, by combining the above results, we can conclude for all $t'\in\mathcal{T}$ we have
\begin{eqnarray*}
   & P_{\mathcal{S}_n}\Bigg( \Big| L_{pop}(\mathcal{M}(\mathcal{S}_n),\mathcal{Z}^{(t')}) - L_{emp}(\mathcal{M}(\mathcal{S}_n),\mathcal{S}_n^{(t')}) \Big| \geq \epsilon(\mathcal{S}^{(t')}_n) + 2B\sqrt{\frac{2K~\ln(2)~+~2~\ln(1/\mathcal{E})}{n^{(t')}}} \Bigg) \\
    &\leq 1 - (1-\mathcal{E})(p_{mr}(\epsilon/K^2))^{K^2}~.
\end{eqnarray*}
\ensuremath{\hfill\blacksquare}

\begin{lemma}
\label{lm: smoothy}

\textbf{(Used for proof of Theorem \ref{th: smoothtau})}
Let $\{\mathcal{S}_n\}_{n=1}^\infty$ be a sequence of nested datasets, each of which includes $n$ i.i.d$.$ samples from $\mu(\mathcal{Z})$, $n=1..\infty$. 
Given a smooth distance metric $\dis_{\mathcal{M}}$, covariate vector $\x$, and $\alpha>0$, 
if there exists a small enough value of ``$a$'' and a large enough value of $N$ 
such that $\mathcal{K}^{(t')}_n(\x) = \{z_k : \dis_\mathcal{M}(\x_k,\x)<a, t_k = t', z_k\in \mathcal{S}_n\}$ is non-empty and  $\alpha > \delta_{\dis_\mathcal{M}}(a)$ for all $n\geq N$
then,
$$P(|E[Y^{(t')}|\x] - \widehat{Y}^{(t')}_{\x}|>\alpha)\leq\exp(-|\mathcal{K}^{(t')}_n(\x)|(\alpha-\delta_{\dis_\mathcal{M}}(a))^2/2\mathbf{C}_y)$$
where $\delta_{\dis_\mathcal{M}}(a)$ is the bound from Definition~\ref{def:smoothdis} (definition of smooth distance metric).
As the above choice of $a$ holds for all $n\geq N$, we have that the bound goes to zero as $n\to\infty$.

\end{lemma}

\noindent \textbf{Proof (Lemma~\ref{lm: smoothy})}.
$\mathcal{K}^{(t')}_n(\x)$ is a matched group of nearest neighbors $z_k$ of unit $z$ such that $d_\mathcal{M}(\x,\mathbf{x}_k)<a$ and treatment indicator $t_k=t'$, i.e.,
\begin{equation*}
    \mathcal{K}^{(t')}_n(\x) = \{z_k \ : \ d_\mathcal{M}(\x,\mathbf{x}_k) < a \text{ and } t_k = t' \}.
\end{equation*}
We estimate the conditional average potential outcome for treatment choice $t'$ and $X=\x$ as $\frac{1}{| \mathcal{K}^{(t')}_n(\x) |} \sum_{z_k\in \mathcal{K}^{(t')}_n(\x) }Y_k$. If $d_\mathcal{M}$ is a smooth distance metric then as all the units in $\mathcal{K}^{(t')}_n(\x)$ have distance to $\x$ less than $a$, for every $z_k$ in $\mathcal{K}^{(t')}_n(\x)$, we have $|\mathbb{E}[Y^{(t')}|X=\x] - \mathbb{E}[Y^{(t')}|X=\x_k]|<\delta_{\dis_\mathcal{M}}(a)$.
Consider $\alpha$ such that $\delta_{\dis_\mathcal{M}}(a)<\alpha$ and $\alpha\leq\mathbf{C}_y$, then
\small
\begin{eqnarray*}
    &P_{\mathcal{S}_n \sim \mu(\mathcal{Z}^n)}\left(\left|\mathbb{E}[Y^{(t')}|X=\x] - \frac{1}{|\mathcal{K}^{(t')}_n(\x)|} \sum_{z_k\in\mathcal{K}^{(t')}_n(\x)}Y_k\right|>\alpha\Bigg| \{\mathbf{X}_i\}, \{T_i\}\right)\\
    &= P\left(\left|\frac{1}{|\mathcal{K}^{(t')}_n(\x)|} \sum_{z_k\in\mathcal{K}^{(t')}_n(\x)}(\mathbb{E}[Y^{(t')}|X=\x] - Y_k)\right|>\alpha\Bigg| \{\mathbf{X}_i\}, \{T_i\}\right).
\end{eqnarray*}
Thus,
\begin{eqnarray*}
    \lefteqn{P\left(\left|\frac{1}{|\mathcal{K}^{(t')}_n(\x)|} \sum_{z_k\in\mathcal{K}^{(t')}_n(\x)}(\mathbb{E}[Y^{(t')}|X=\x] - Y_k)\right|>\alpha\Bigg| \{\mathbf{X}_i\}, \{T_i\}\right)}\\
    &=& 
    P\left(\left|\sum_{z_k\in\mathcal{K}^{(t')}_n(\x)} \mathbb{E}[Y^{(t')}|X=\x] - \mathbb{E}[Y^{(t')}|X=\mathbf{X}_k] + \mathbb{E}[Y^{(t')}|X=\mathbf{X}_k] - Y_k\right|>|\mathcal{K}|\alpha\Bigg| \{\mathbf{X}_i\}, \{T_i\}\right)\\
    && \\
    && \textrm{and using the triangle inequality, }\\
&\leq& 
    P\Bigg(\sum_{z_k\in\mathcal{K}^{(t')}_n(\x)} \left|\mathbb{E}[Y^{(t')}|X=\x] - \mathbb{E}[Y^{(t')}|X=\mathbf{X}_k]\right| +
    \\
    && \hspace{2cm} \left|\sum_{z_k\in\mathcal{K}^{(t')}_n(\x)} \mathbb{E}[Y^{(t')}|X=\mathbf{X}_k] - Y_k\right|>|\mathcal{K}^{(t')}_n(\x)|\alpha\Bigg| \{\mathbf{X}_i\}, \{T_i\}\Bigg). \\
    && \\
    && \textrm{ By the definition of smooth distance metric, }\\
   &\leq& P\left(\sum_{z_k\in\mathcal{K}^{(t')}_n(\x)}\delta_{\dis_\mathcal{M}}(a)  + \left|\sum_{z_k\in\mathcal{K}^{(t')}_n(\x)} \mathbb{E}[Y^{(t')}|X=\mathbf{X}_k] - Y_k\right|>|\mathcal{K}^{(t')}_n(\x)|\alpha\Bigg| \{\mathbf{X}_i\}, \{T_i\}\right)\\
    &=&
    P\left(\left|\sum_{z_k\in\mathcal{K}^{(t')}_n(\x)} \mathbb{E}[Y^{(t')}|X=\mathbf{X}_k] - Y_k\right|>|\mathcal{K}^{(t')}_n(\x)|(\alpha - \delta_{\dis_\mathcal{M}}(a) ) \Bigg| \{\mathbf{X}_i\}, \{T_i\} \right).\\
    &&\\
    && \textrm{and by Hoeffding's inequality, }\\
    & \leq& 
    2\exp\left(\frac{-2 |\mathcal{K}^{(t')}_n(\x)|(\alpha-\delta_{\dis_\mathcal{M}}(a))^2}{4\mathbf{C}_y}\right)=
    2\exp\left(\frac{- |\mathcal{K}^{(t')}_n(\x)|(\alpha-\delta_{\dis_\mathcal{M}}(a))^2}{2\mathbf{C}_y}\right)\;.
\end{eqnarray*}
\normalsize
Thus,
\begin{eqnarray*}
     && P\left(\left|\mathbb{E}[Y^{(t')}|X=\x] - \frac{1}{|\mathcal{K}^{(t')}_n(\x)|}\sum_{z_k\in\mathcal{K}^{(t')}_n(\x)}Y_k\right|>\alpha\Bigg| \{\mathbf{X}_i\}, \{T_i\}\right) \\ && \;\;\; \leq 2\exp\left(\frac{- |\mathcal{K}^{(t')}_n(\x)|(\alpha-\delta_{\dis_\mathcal{M}}(a))^2}{2\mathbf{C}_y}\right).
\end{eqnarray*}
As $n\rightarrow \infty$, for a constant $a$ (and hence constant $\delta_{\dis_\mathcal{M}}(a)$), a constant $\alpha$, and letting the number of units matched to the target unit go to infinity: $|\mathcal{K}^{(t')}_n(\x)|\rightarrow\infty$, we have $$\text{lim}_{n\rightarrow \infty} P\left(|\mathbb{E}[Y^{(t')}|X=\x] - \frac{1}{|\mathcal{K}^{(t')}_n(\x)|}\sum_{z_k\in\mathcal{K}^{(t')}_n(\x)}Y_k|>\alpha\Bigg| \{\mathbf{X}_i\}, \{T_i\}\right) = 0.$$
\ensuremath{\hfill\blacksquare}
\begin{lemma}
\label{lm: ytotau}
\textbf{(Also used for proof of Theorem \ref{th: smoothtau})}
If we can estimate the conditional average potential outcomes using a finite sample $\mathcal{S}_n\overset{i.i.d}{\sim}\mu(\mathcal{Z}^n)$ such that for all $t'$, $\mathbb{E}[Y^{(t')}|X=\x]$ and the estimate, $\hat{Y}^{(t')}_\x$ are farther than $\epsilon'$ with probability less than $\delta'(\epsilon',n)$ for any given $z=(\x,y,t)\in\mathcal{Z}$ and $t \in \mathcal{T}$,
then the estimated conditional average treatment effect $\hat{\tau}(\x)$ using a finite sample $\mathcal{S}_n\overset{i.i.d}{\sim}\mu(\mathcal{Z}^n)$ and the true conditional average treatment effect $\tau(\x)$ are farther than $\epsilon$ with probability less than $2\delta'(\frac{\epsilon}{2},n)$.
\begin{eqnarray*} 
\forall t\in \mathcal{T}, &&P_{\mathcal{S}_n \sim \mu(\mathcal{Z}^n)}\Big( |\hat{Y}^{(t')}_\x - \mathbb{E}[Y^{(t')}|X=\x]| \geq \epsilon' \Big) \leq \delta'(\epsilon',n) \\ && \implies P_{\mathcal{S}_n \sim \mu(\mathcal{Z}^n)}\Big( |\hat{\tau}(\x) - \tau(\x)| \geq \epsilon \Big) \leq 2\delta'\Big(\frac{\epsilon}{2},n\Big). \end{eqnarray*} 
\end{lemma}
\textbf{Proof (Lemma~\ref{lm: ytotau})}.
We are given in the statement that for any $\epsilon'>0$, we can find a $ \delta^{'}(\epsilon',n)$ such that we can estimate outcomes well, i.e.,
\begin{equation*}
    \forall z\in\mathcal{Z},~\forall t'\in \mathcal{T},~P_{\mathcal{S}_n \sim \mu(\mathcal{Z}^n)}(|\hat{Y}^{(t')}_\x - \mathbb{E}[Y^{(t')}|X=\x]|\geq \epsilon^{'})\leq \delta^{'}(\epsilon',n).
\end{equation*}
We can further deduce from the union bound that
\begin{equation}
    \label{eq: andeq}
    P\Bigg( \bigvee_{t'\in\mathcal{T}}~\Big(|\hat{Y}^{(t')}_\x - \mathbb{E}[Y^{(t')}|X=\x]| \geq \epsilon^{'}\Big) \Bigg) \leq |\mathcal{T}|~\delta^{'}(\epsilon',n).
\end{equation}
By the triangle inequality, we also know that
\begin{equation}
\label{eq: treq1}
  \sum_{t'\in\mathcal{T}}\Big|\hat{Y}^{(t')}_\x - \mathbb{E}[Y^{(t')}|X=\x]\Big|
    \geq \Bigg| \sum_{t'\in\mathcal{T}}\Big(\hat{Y}^{(t')}_\x - \mathbb{E}[Y^{(t')}|X=\x]\Big) \Bigg|.
\end{equation}
From Equation~\ref{eq: andeq}, we have
\begin{equation*}
    P\Bigg( \sum_{t'\in\mathcal{T}}~\Big(|\hat{Y}^{(t')}_\x - \mathbb{E}[Y^{(t')}|X=\x]|\Big)  \geq |\mathcal{T}|\epsilon^{'}\Bigg) \leq |\mathcal{T}|~\delta^{'}(\epsilon',n).
\end{equation*}
Applying the triangle inequality from Equation~\ref{eq: treq1},
\begin{equation*}
    P\Bigg( \Big| \sum_{t'\in\mathcal{T}}~\Big(\hat{Y}^{(t')}_\x - \mathbb{E}[Y^{(t')}|X=\x]\Big)\Big|  \geq |\mathcal{T}|\epsilon^{'}\Bigg) \leq |\mathcal{T}|~\delta^{'}(\epsilon',n).
\end{equation*}
Considering the case where $\mathcal{T}=\{0,1\}$,
\begin{equation*}
    P\Bigg( \Big| \hat{\tau}(\x) - \tau(\x)\Big|  \geq 2\epsilon^{'}\Bigg) \leq 2\delta^{'}(\epsilon',n).
\end{equation*}
Hence, we can conclude that
\begin{equation*}
    P\Bigg( \Big| \hat{\tau}(\x) - \tau(\x)\Big|  \geq \epsilon\Bigg) \leq 2\delta^{'}\Big(\frac{\epsilon}{2},n\Big).
\end{equation*}
\ensuremath{\hfill\blacksquare}

\noindent \textbf{Proof (Theorem~\ref{th: smoothtau})}.
The proof of Theorem \ref{th: smoothtau} follows directly by substituting the result of Lemma \ref{lm: smoothy} into Lemma \ref{lm: ytotau}.

\ensuremath{\hfill\blacksquare}

\noindent \textbf{Proof (Theorem~\ref{th: gen})}. By Theorem~\ref{th: robust}, we know that the distance metric $ \mathcal{M}(\cdot) $ learned using \textsc{MALTS} is $(\mathbf{N}(\gamma,\mathcal{X},\|\cdot\|_2),\beta)$-multirobust with probability more than $1-\exp\left(-\frac{\beta^2 \left(\rho^{(t')}_{\gamma}\right)^2}{ n^{(t')} B^2 }\right)$.
Also, inferring from Lemma~\ref{lm: whpavgloss}, for any arbitrary $\forall t' \in \mathcal{T}$ and $\mathcal{E}>0$ we have that with probability at least $$(1-\mathcal{E})\left(1-\exp\left(-\frac{\beta^2 \left(\rho^{(t')}_{\gamma}\right)^2}{ K^2 n^{(t')} B^2 }\right)\right)^{K^2}$$ 
with respect to the random draw of data to form $\mathcal{S}_n$, we have
\begin{eqnarray*}
\Big| L_{pop}(\mathcal{M}(\mathcal{S}_n),\mathcal{Z}^{(t')}) - L_{emp}(\mathcal{M}(\mathcal{S}_n),\mathcal{S}_n^{(t')}) \Big| \leq \beta + 2B\sqrt{\frac{2K~\ln(2)~+~2~\ln(1/\mathcal{E})}{n^{(t')}}}. 
\end{eqnarray*}

Let $\rho_\gamma = \min_{t'} \rho^{(t')}_\gamma$. Then, summing over all possible $t'\in\mathcal{T}$ we have that with probability at least:
$$(1-\mathcal{E})^{|\mathcal{T}|}\left(1-\exp\left(-\frac{\beta^2 \left(\rho^{(t')}_{\gamma}\right)^2}{ K^2 n^{(t')} B^2 }\right)\right)^{|\mathcal{T}|K^2}$$
with respect to the random draw of data to form $\mathcal{S}_n$, we have:
\begin{eqnarray*}
\sum_{t'\in\mathcal{T}} \Big| L_{pop}(\mathcal{M}(\mathcal{S}_n),\mathcal{Z}^{(t')}) - L_{emp}(\mathcal{M}(\mathcal{S}_n),\mathcal{S}_n^{(t')}) \Big| \leq
     2|\mathcal{T}|\beta+ \sum_{t'\in\mathcal{T}} 2B\sqrt{\frac{2K~\ln(2)~+~2~\ln(1/\mathcal{E})}{n^{(t')}}}.
\end{eqnarray*}
$\gamma$ in Theorem~\ref{th: robust} was arbitrary, allowing us to take it to 0 in such a way that $K$ increases at a rate smaller than $\min_{t'} n^{(t')}$ increases, and $\rho_\gamma$ increases at rate faster than or equal to $O(n)$ to $\infty$ as $n$ approaches $\infty$. Thus we can reduce $\beta^2$ to $0$ at a rate slower than $\frac{1}{\rho^2_\gamma}$. $\mathcal{E}$ was also set arbitrarily, allowing us to take it to 0 slowly enough such that as $n\to\infty$, each of the $n ^{(t')}\to\infty$ and we thus have: 
\begin{equation}
    \lim_{n\to\infty} \Bigg( \sum_{t'\in\mathcal{T}} \Big| L_{pop}(\mathcal{M}(\mathcal{S}_{tr}),\mathcal{Z}^{(t')}) - L_{emp}(\mathcal{M}(\mathcal{S}_{tr}),\mathcal{S}^{(t')}_{tr}) \Big| \Bigg) = 0.
\end{equation}
\ensuremath{\hfill\blacksquare}

     \section{}\label{sec:appendixB}
\label{sec:var_overlap}

\subsection*{Limited Overlap and Performance}
\begin{figure}
    \centering
    \includegraphics[width = 0.8\textwidth]{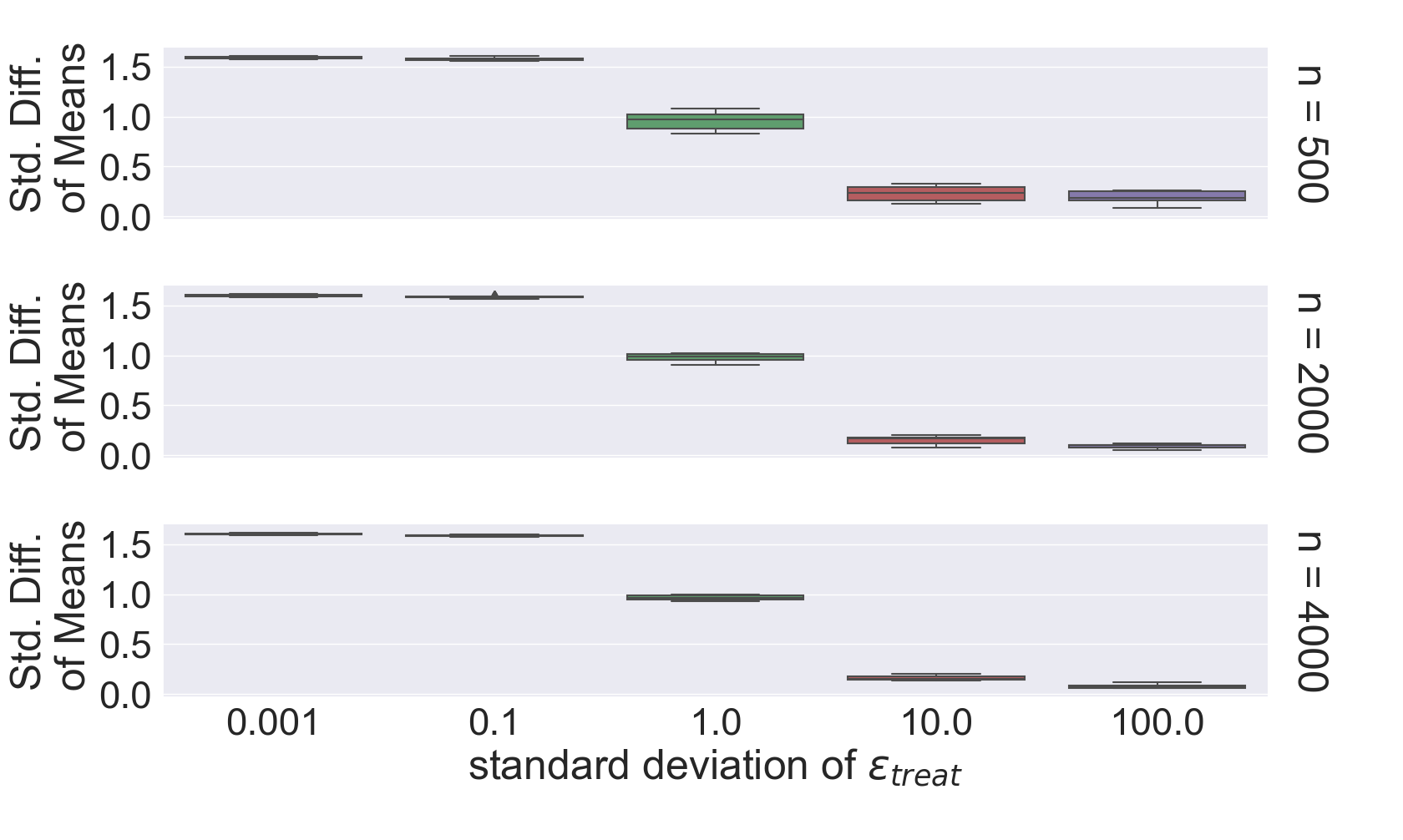}
    \caption{Standardized difference of means between covariates of treated and control units decreases as $\text{std}(\epsilon_{\textrm{treat}})$ increases. We increase the value of $\text{std}(\epsilon_{\textrm{treat}})$ in the DGP for treatment allocation which increases the overlap in the treated and control groups. We generate data with $p$ equal to 20 covariates and for values of $n\in\{500,2000,4000\}$.}
    \label{fig:overlap_r}
\end{figure}

\begin{figure}
        \centering
        \includegraphics[width = 0.85\textwidth]{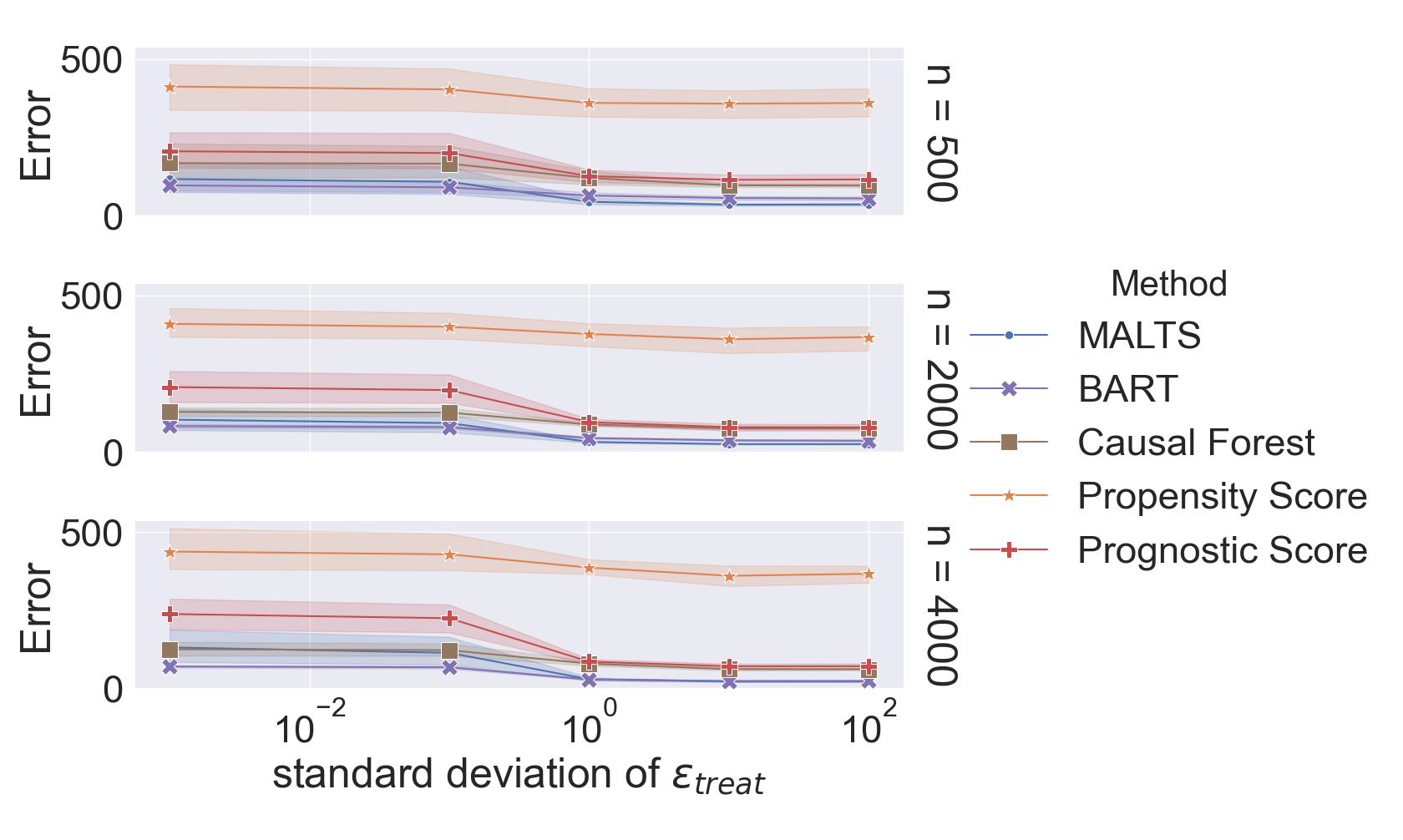}\\
        (a)\\
        \includegraphics[width =  0.85\textwidth]{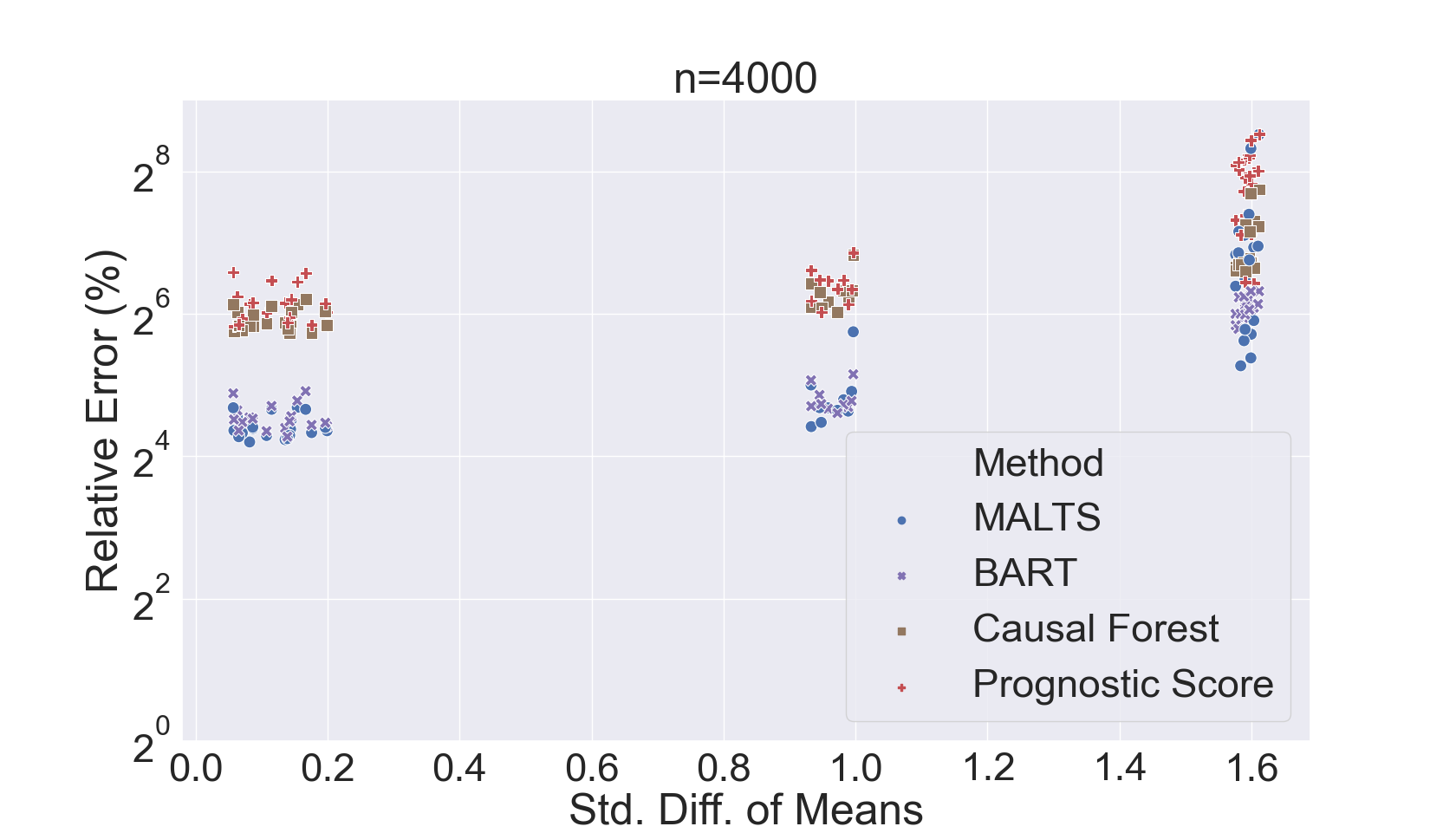}\\
        (b)
        \caption{(a) Trend plot comparison of MALTS performance measured as mean relative error for CATE estimation under different levels of overlap, measured as a function of standard deviation of $\epsilon_{\textrm{treat}}$ (the scale of noise in treatment allocation process). Higher values of the standard devation of $\epsilon_{\textrm{treat}}$ corresponds to more overlap between the control and the treated groups. (b) Scatterplot comparing MALTS' performance, as mean CATE estimation error, under different levels of overlap. Overlap is measured as standardized difference of means for $n=4000$. Larger values of standardized difference of means corresponds to less overlap between the control and treated groups.}
        \label{fig:overlap}
    \end{figure}

We use the DGP described in Section~\ref{dgp1} with 2 relevant continuous covariates and no irrelevant covariates for the limited overlap experiments. Further, we set the parameters of the DGP as follows:
$\mu = 1$, $\Sigma = 1.5$, $\phi=0.5$, and $c = 2$. We performed experiments on the overlap by changing the standard deviation of the noise term $\epsilon_{\textrm{treat}}$ from $0.001$ to $100$ in the treatment assignment equation of the DGP and measured CATE estimation error for MALTS in comparison with other methods for each of the scenarios. A lower variance leads to small overlap, i.e., large standardized difference of means, whereas large variance creates small standardized difference of means and high overlap.
Figure~\ref{fig:overlap_r} shows the relationship of $\epsilon_{\textrm{treat}}$ and standardized difference of means between the covariate set of treated and control units. Figure~\ref{fig:overlap} shows the performance of MALTS in comparison with other methods in predicting CATEs for multiple dataset sizes -- $n\in\{500,2000, 4000\}$ -- with $p=20$, and different levels of overlap. MALTS' performance is largely insensitive to limited overlap, however, the performance deteriorates if the control and the treated units are very different. The primary reason for MALTS deterioration of performance under almost no-overlap compared to BART is because matching methods like MALTS can be conceptualized as \textit{interpolation}, unlike regression approaches that explicitly models the potential outcomes' surfaces, which is closer to \textit{extrapolation}.

    
\subsection*{Sensitivity Analysis}
\begin{figure}
        \centering
        \includegraphics[width=0.85\textwidth]{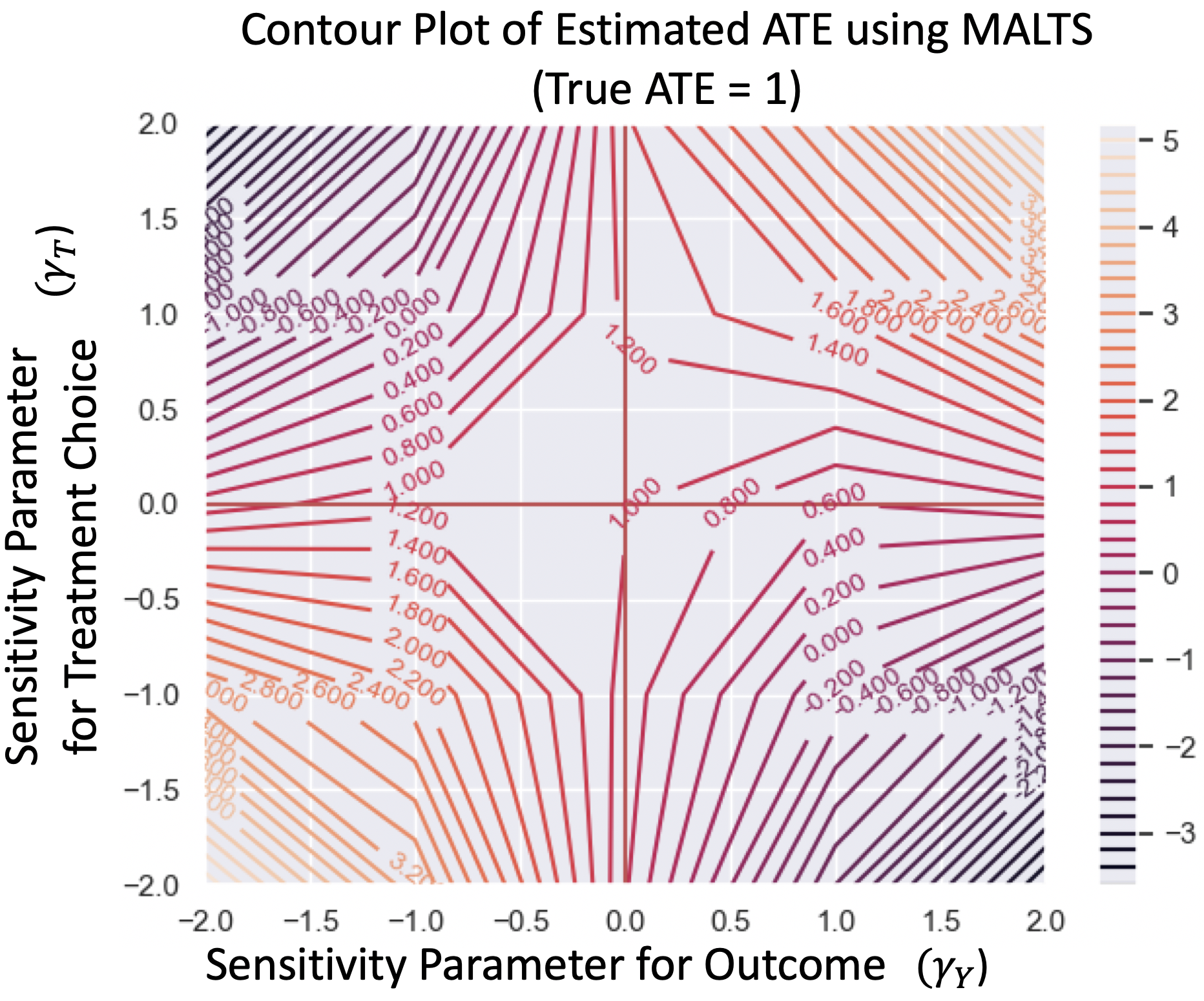}
        \caption{Sensitivity analysis contour plot of the ATE estimation using MALTS. The best performance is when there is no unmeasured confounding, which is when both sensitivity parameters are 0.
        When the sensitivity parameter for the outcome is 2, the unmeasured confounder is approximately as important as the total of the two known covariates and performance degrades.}
        \label{fig:sensitivity}
    \end{figure}

We performed a \textit{sensitivity analysis} of MALTS on a data generation setup with a constant treatment effect, two observed relevant covariates, and an unobserved confounder affecting the probability distributions of outcome as well as the choice of treatment. The unobserved confounder has a linear relationship with the outcome, with the value of the coefficient equal to the ``sensitivity parameter of the outcome'' ($\gamma_Y$) and the choice of treatment with the value of the coefficient equal to the ``sensitivity parameter of the treatment'' ($\gamma_T$).
\begin{eqnarray*}
    && x_{i,1},x_{i,2}, u_i \overset{iid}{\sim} \mathcal{N}(0,1)\\
    && \epsilon_{i,0}, \epsilon_{i,1} \overset{iid}{\sim} \mathcal{N}(0,1)\\
    y_i^{(0)} &=& x_{i,1} + x_{i,2} + \gamma_Y u_i + \epsilon_{i,0} \\
    y_i^{(1)} &=& x_{i,1} + x_{i,2} + \gamma_Y u_i + 1 + \epsilon_{i,1} \\
    t_i &=& \text{Bernoulli}\left(\text{expit}\left( x_{i,1} + x_{i,2} + \gamma_T u_i - 2 \right) \right)\\
    y_i &=& t_iy_i^{(1)} + (1-t_i)y_i^{(0)}
\end{eqnarray*}
Figure~\ref{fig:sensitivity} shows the contour plot of ATE estimates produced by MALTS as we change the sensitivity parameters in the data generation process. Here, the true ATE is equal to 1. The plot indicates that as long as the unmeasured confounders are approximately half as important as either of the two observed covariates ($\gamma_T$ or $\gamma_Y$ is below 0.5), MALTS performance is stable. At the extreme, which is when the unobserved confounder is as important as the total of the two observed covariates ($\gamma_T$ or $\gamma_Y$ is 2), the performance (expectedly) degrades.
     \section{}\label{sec:appendixC}
In this section, we discuss our implementation of existing causal inference methods like genmatch, propensity score matching, BART, causal forest, difference of random forest,  prognostic score matching and FLAME. In Section~\ref{sec:Experiments}, we compare the performance of each of these methods with \malts.

We used MatchIt's implementation of genmatch and propensity score matching as it is commonly used by empiricists \citep{matchit2011}. We allowed matching with replacement for creating match groups and estimating CATEs. As MatchIt returns only match groups and CATE estimates for treated units (and not control units), then in order to estimate CATEs for control units we flipped the sign of the treatment indicators and estimated negative CATEs (we have to estimate negative CATE in this case because we flipped the sign of the treatment indicator causing CATE estimates to become negative CATE estimates). We merged the CATE estimates for the treated units and control units to get the CATE estimates for every unit in the dataset.

We used the causal forest algorithm as implemented in the `grf' package in R. The settings for causal forest were set to the default designed by the `grf' developer with number of trees equal to $2000$ and $\sqrt{p}+20$ variables tried for each split. 

We performed the same 5-fold CATE estimation procedure for causal forest, analogous to the one used for estimating CATEs using MALTS. We estimated CATEs for both the treated and control units in each estimation set. 

We used Vincent Dorie's R implementation of BART \citep{dbart}. We performed the same 5-fold CATE estimation using BART that we used for MALTS. For each of the $\eta$ folds, we trained two BART models, one for learning the response function for estimating the potential outcome under control and the other response function for estimating potential outcome under treatment using the training set. The CATEs were estimated by taking the difference of estimated response functions of treated and control units in the estimation set. 

We also implemented a 5-fold FLAME CATE estimation procedure analogous to the one used by MALTS.

Lastly, we implemented 5-fold prognostic score matching using a random forest approach to model the prognostic score function. We fit a model for control units and a model for treated units using the data in the training set. To estimate the CATE for a treated unit in the estimation set, we found k-nearest neighbors in the control set with a similar estimated prognostic score. Analogously, we estimated the CATEs for the control units in the estimation set using the k-nearest treated units with similarity measured using the prognostic score.

\section{}\label{sec:appendixD}
In this section, we show expanded matched groups (including all treatment and control units, not just control units) using propensity score matching, prognostic score matching and MALTS for unit id-1 from Table~\ref{tb:ex_mg_1}. MALTS and prognostic score matching are implemented as described in \ref{sec:appendixC} with $K=10$ nearest neighbors, where $K$ was selected by cross-validation for MALTS and used also for the other methods. Propensity score matching was implemented using the MatchIt package and operationalized with 10 nearest neighbor matching. 

Here, MALTS produces a high quality matched group, as shown in Table~\ref{tab:match-group-example}. While MALTS matches units based on a learned distance metric over the covariate space, prognostic score matching matches units based on a single prognostic value and propensity score matching matches units based on a single propensity value. 
We note that there is only one matched unit in common between the matched groups from prognostic score matching and MALTS (unit 116) and there are two units in common between the matched groups from propensity score matching and MALTS (units 330 and 416); the matched groups are almost entirely different between the three methods.

\begin{table}
\centering
\caption{Example Matched Group using (a) our approach, (b) prognostic score \citep{hansen2008prognostic}, and (c) propensity score matching \citep{rosenbaum1983central} for a query unit in the Lalonde dataset (top rows). It matched closely on almost all covariates such as age, education, marital status, whether the person had an academic degree, and income in 1975. In contrast, prognostic and propensity scores did not match closely on factors such as education, age, marital status and income. Bold is used to denote disagreement between the query unit and its matched group.
}
\label{tb:ex_mg_1}
\resizebox{\textwidth}{!}{\begin{tabular}{l|c|rrrrrrr|r}
\hline
\multicolumn{1}{c}{} & \multicolumn{1}{c|}{\textbf{Treatment}} & \multicolumn{7}{c|}{\textbf{Covariates}} & \multicolumn{1}{c}{\textbf{Outcome}} \\ \hline
\multicolumn{1}{l|}{\textit{\textbf{Unit ID}}} & \multicolumn{1}{r|}{\textbf{Treated}} & \textbf{Age} & \textbf{Education} & \textbf{Black} & \textbf{Hispanic} & \textbf{Married} & \textbf{No-Degree} & \multicolumn{1}{r|}{\textbf{Income-1975}} & \textbf{Income-1978} \\ \hline
Query: 1 & Yes & 22 & 9 & No & Yes & No & Yes & \$0 & \$3596 \\ \hline
\multicolumn{10}{c}{\textbf{(a) Our Approach (MALTS)}} \\ \hline
94 & Yes & 23 & 8 & No & Yes & No & Yes & \$0 & \$3881 \\
330 & No & 22 & 8 & No & Yes & No & Yes & \$0 & \$9921 \\
299 & No & 22 & 9 & \textbf{Yes} & \textbf{No} & No & Yes & \$0 & \$0 \\
5 & Yes & 22 & 9 & \textbf{Yes} & \textbf{No} & No & Yes & \$0 & \$4056 \\
82 & Yes & 21 & 9 & \textbf{Yes} & \textbf{No} & No & Yes & \$0 & \$0 \\
416 & No & 22 & 9 & \textbf{Yes} & \textbf{No} & No & Yes & \$0 & \$12898 \\
333 & No & 21 & 9 & \textbf{Yes} & \textbf{No} & No & Yes & \$0 & \$3343 \\
292 & Yes & 20 & 9 & \textbf{Yes} & \textbf{No} & No & Yes & \$0 & \$8882 \\
17 & Yes & 23 & 10 & \textbf{Yes} & \textbf{No} & No & Yes & \$0 & \$7693 \\
116 & Yes & 24 & 10 & \textbf{Yes} & \textbf{No} & No & Yes & \$0 & \$0 \\ \hline
\multicolumn{10}{c}{\textbf{(b) Prognostic Scores}} \\ \hline
154 & Yes & 22 & 10 & \textbf{Yes} & \textbf{No} & No & Yes & \textbf{\$1071} & \$7315 \\
56 & Yes & \textbf{30} & \textbf{11} & \textbf{Yes} & \textbf{No} & \textbf{Yes} & Yes & \$0 & \$591 \\
100 & Yes & \textbf{17} & 10 & \textbf{Yes} & \textbf{No} & No & Yes & \$0 & \$0 \\
109 & Yes & 18 & 9 & \textbf{Yes} & \textbf{No} & No & Yes & \$0 & \$4483 \\
141 & Yes & 25 & 8 & \textbf{Yes} & \textbf{No} & No & Yes & \textbf{\$37432} & \$2347 \\
286 & Yes & 23 & \textbf{12} & No & Yes & No & \textbf{No} & \textbf{\$1117} & \$559 \\
338 & No & \textbf{44} & 9 & \textbf{Yes} & \textbf{No} & No & Yes & \$0 & \$9722 \\
340 & No & 22 & \textbf{12} & \textbf{Yes} & \textbf{No} & No & \textbf{No} & \textbf{\$532} & \$1333 \\
355 & No & 18 & 10 & No & Yes & No & Yes & \$0 & \$1859 \\
116 & Yes & 24 & 10 & \textbf{Yes} & \textbf{No} & No & Yes & \$0 & \$0 \\ \hline
\multicolumn{10}{c}{\textbf{(c) Propensity Scores}} \\ \hline
416 & No & 22 & 9 & \textbf{Yes} & \textbf{No} & No & Yes & \$0 & \$12898 \\
451 & No & 22 & 8 & \textbf{Yes} & \textbf{No} & No & Yes & \$0 & \$1391 \\
330 & No & 22 & 8 & No & Yes & No & Yes & \$0 & \$9921 \\
407 & No & 20 & \textbf{12} & \textbf{Yes} & \textbf{No} & No & \textbf{No} & \textbf{\$1371} & \$20893 \\
626 & No & 18 & 10 & \textbf{Yes} & \textbf{No} & No & Yes & \textbf{\$2682} & \$0 \\
774 & No & 21 & \textbf{13} & No & \textbf{No} & No & \textbf{No} & \textbf{\$693} & \$2660 \\
402 & No & 22 & \textbf{11} & \textbf{Yes} & \textbf{No} & \textbf{Yes} & Yes & \$0 & \$1698 \\
925 & No & 21 & \textbf{12} & \textbf{Yes} & \textbf{No} & No & \textbf{No} & \textbf{\$716} & \$22166 \\
879 & No & 22 & \textbf{12} & \textbf{Yes} & \textbf{No} & \textbf{Yes} & \textbf{No} & \$0 & \$665 \\
788 & No & 22 & \textbf{12} & \textbf{Yes} & \textbf{No} & \textbf{Yes} & \textbf{No} & \$0 & \$0 \\ \hline
\end{tabular}}
\end{table}

\section{}\label{sec:appendixE}
\begin{figure}
        \centering
        \includegraphics[width=0.85\textwidth]{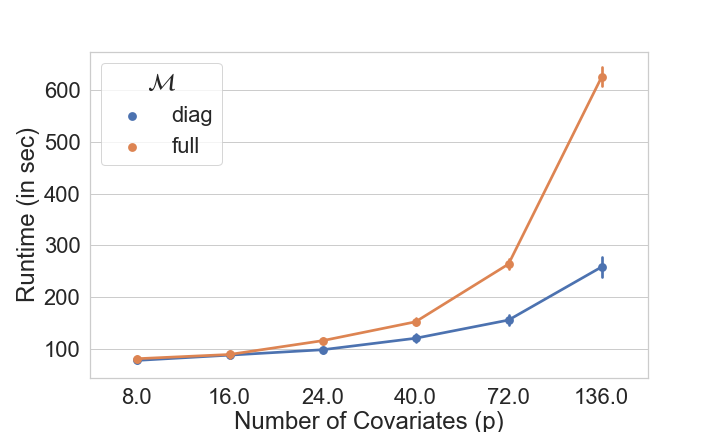}
        \caption{Run time for MALTS when distance metric is constrained to have diagonal $\mathcal{M}$, compared with distance metric where $\mathcal{M}$ is a full-rank positive semi-definite matrix.}
        \label{fig:timings}
    \end{figure}

In this section, we study the computation time required by MALTS to learn an optimal distance metric and estimate CATEs using matching. We study this by increasing the number of covariates from 8 to 136 (where the number of relevant covariates is 8 and others are irrelevant) keeping the number of samples constant at 2048. We compare the performance of MALTS for unrestricted stretch matrices (referred as `full $\mathcal{M}$') with a case when the stretch matrix is restricted to diagonal matrices (referred as `diagonal $\mathcal{M}$'). Figure~\ref{fig:timings} shows that the difference in the computational time between the two cases increases approximately quadratically as the number of covariates increases. This is because inverting full $\mathcal{M}$ is $O(p^2)$ more costly than inverting a diagonal $\mathcal{M}$.
\end{document}